\documentclass[reprint, amsmath,amssymb, aps, twocolumn]{revtex4-2}
\usepackage{graphicx}% Include figure files
\usepackage{dcolumn}% Align table columns on decimal point
\usepackage{bm}% bold math
\usepackage{blindtext}
\usepackage{amsmath}
\usepackage{xcolor}

\usepackage{subcaption}

\begin{document}

\preprint{APS/123-QED}

\title{Polaritonic quantisation in nonlocal polar materials}% Force line breaks with \\

\author{Christopher R. Gubbin}
\author{Simone De Liberato}%
 \email{s.de-liberato@soton.ac.uk}
\affiliation{%
 Department of Physics and Astronomy\\
 University of Southampton.
}%

\date{\today}% It is always \today, today,

\begin{abstract}
In the Reststrahlen region, between the transverse and longitudinal phonon frequencies, polar dielectric materials respond metallically to light and the resulting strong light-matter interactions can lead to the formation of hybrid quasiparticles termed surface phonon polaritons. Recent works have demonstrated that when an optical system contains nanoscale polar elements these excitations can acquire a longitudinal field component as a result of the material dispersion of the lattice, leading to the formation of secondary quasiparticles termed longitudinal-transverse polaritons. In this work we build on previous macroscopic electromagnetic theories developing a full second-quantised theory of longitudinal-transverse polaritons. Beginning from the Hamiltonian of the light-matter system we treat distortion to the lattice introducing an elastic free energy. We then diagonalise the Hamiltonian, demonstrating the equations of motion for the polariton are equivalent to the those of macroscopic electromagnetism and quantise the nonlocal operators. Finally we demonstrate how to reconstruct the electromagnetic fields in terms of the polariton states and explore polariton induced enhancements of the  Purcell factor. These results demonstrate how nonlocality can narrow, enhance and spectrally tune near field emission with applications in mid-infrared sensing.
\end{abstract}

\maketitle

\section*{Introduction}
Surface phonon polaritons (SPhPs) are hybrid light-matter excitations, formed when a photon interacts with the optical phonon modes of a polar lattice. Like plasmons in the visible spectral region they allow for miniaturisation of photonic resonators below the diffraction limit \cite{Caldwell2013}, with applications in mid-infrared sensing \cite{Berte2018}, nonlinear optics \cite{Gubbin2017c, Razdolski2018, Kitade2021} and the design of thermal emitters \cite{Greffet2002, Schuller2008, Arnold2012, Lu2021} .\\
In a simple model of polar optics the lattice is considered to be non-dispersive, and is described by a frequency-dependant dielectric function parameterised by the optical phonon frequencies at zone-centre. This model works extremely well in regimes where the dispersion of the optic phonon branches can be neglected, however when the system approaches the nanoscale this is no longer necessarily the case. Systems where material dispersion is important are termed optically nonlocal, meaning that the applied field at one point can affect the response at another. This is a consequence of energy transport in the matter, for example through bulk plasma waves in nano-plasmonic systems \cite{Ciraci2013}. This transfer of energy from the photon field leads to enhanced damping, spectral shifts and ultimately places an upper-limit on how tightly the field can be confined \cite{Ciraci2012, Fernandez-Dominguez2012, Mortensen2014, Luo2013}. \\
Nonlocal effects in polar systems are fundamentally different to those in plasmonic systems because of the opposite dispersion of bulk optical phonon and plasma waves. In the Reststrahlen region where the dielectric function is negative and SPhPs exist, polar crystals also support propagative LO phonon excitations, allowing for resonant coupling between discrete LO phonon modes and SPhPs. These nonlocal effects in polar systems were first demonstrated in a recent work which showed strong coupling between localised SPhPs in a 4H-SiC nanopillar and zone-folded optical phonons \cite{Gubbin2019}. The resulting excitations are hybrid modes termed longitudinal-transverse polaritons (LTPs), have a transverse electric field from their photonic component and a longitudinal one from the phonon field. This unique property has led to LTPs being proposed as a platform for mid-infrared optoelectronics as a result of the possibility of exciting them through longitudinal electrical currents, while still outcoupling to transverse free-space radiation in the far-field \cite{Gubbin2021b}.\\
A recent series of publications has studied LTPs in more general systems, demonstrating them to be a general feature of polar resonators at the nanoscale \cite{Gubbin2020, Gubbin2020b, Gubbin2020c} and a similar phenomenology has also recently been observed in a Yukawa fluid \cite{Yakovlev2020}. These works follow the approach of nonlocal plasmonics, starting from the macroscopic Maxwell equations, introducing new macroscopic fields to describe phonon modes in the lattice and matching fields at material boundaries considering the flow of energy in the system. They have been utilised to explain the emergence of anomalous modes in complex crystal hybrid structures, macroscopic systems comprised of hundreds of nanoscopic polar layers \cite{Ratchford2019, Gubbin2020} and have proved able to calculate the electromagnetic response.\\
Polariton systems are typically described in second-quantisation formalisms \cite{Hopfield1958, Alpeggiani2014, Archambault2010, Gubbin2016b} which are attractive because they allow for a transparent understanding of energy distribution between the different light and matter excitations from which the polariton is composed, and for calculation of nonlinear polariton-polariton \cite{Gubbin2017b, Carusotto2013, Tan2020} or electron-polariton scattering \cite{Efimkin2021}. Although recently a second-quantisation theory of LTPs was presented, this was based on a demonstration of equivalence between Maxwell's equations and a model Hamiltonian in a planar waveguide \cite{Gubbin2021} and is not suitable for extension to describe the complex modal geometries typically utilised in SPhP optics.\\
In this work we develop a full quantum theory of LTPs. Starting from the Hamiltonian of a spectrally dispersive, spatially inhomogeneous polar dielectric we introduce a new free energy to account for the elastic energy of optical phonons. We then diagonalise this Hamiltonian, demonstrating that the resulting equations of motion for the Hopfield fields are equivalent to the nonlocal Maxwell equations \cite{Gubbin2020}. Operators for the physical fields are reconstructred from the Hopfield fields, the boundary conditions satisfied by those fields are determined and the quantisation conditions for the polaritonic modes are derived. Finally we utilise our theory to demonstrate the nonlocal emission from dipoles embedded near a polar nanolayer.

\section{Theory}
Our starting point is the Hamiltonian density in a local, non-magnetic polar dielectric including spectral dispersion. We consider the material to be piecewise homogeneous such that we may employ Fourier analysis in each region. The system's quantum Hamiltonian is given by
\begin{align}
	\hat{\mathcal{H}}_0 = \frac{\hat{\mathrm{D}}^2}{2 \epsilon_{0}\epsilon_{\infty}}  + \frac{\mu_0 \hat{\mathrm{H}}^2 }{2} + \frac{\hat{\mathrm{Q}}^2}{2 \rho} + \frac{\rho \omega_{\mathrm{L}}^2 \hat{\mathrm{X}}^2}{2} - \frac{\kappa \hat{\mathbf{D}} \cdot \hat{\mathbf{X}}}{\epsilon_0 \epsilon_{\infty}} , \label{eq:Hamloc}
\end{align}
where $\hat{\mathrm{D}} \; (\hat{\mathrm{H}})$ is the electric displacement (magnetic) field operator, $\hat{\mathrm{X}}$ is the ionic displacement operator, $\hat{\mathrm{Q}}$ is the momentum operator, $\omega_{\mathrm{L}}$ is the longitudinal optical phonon frequency at zone-centre, $\epsilon_{\infty}$ is the high-frequency dielectric constant and $\rho$ is the crystal mass density \cite{Gubbin2016b}. The oscillator strength $\kappa$ characterises the width of the Reststrahlen that relates to the phonon frequencies through
\begin{equation}
	\frac{\kappa^2}{\rho \epsilon_0 \epsilon_{\infty}} = \omega_{\mathrm{L}}^2 - \omega_{\mathrm{T}}^2 = \omega_{\mathrm{L}}^2 \left[ 1 - \frac{\epsilon_{\infty}}{\epsilon_{\mathrm{st}}}\right], \label{eq:LST}
\end{equation}
where in the latter equality we utilised the Lydanne-Sachs-Teller relation, linking the crystal's transverse and longitudinal optic phonon frequencies ($\omega_{\mathrm{T}}$ and $\omega_{\mathrm{L}}$) using the static dielectric function of the lattice $\epsilon_{\mathrm{st}}$ \cite{Lyddane1941}.
The Hamiltonian Eq.~\ref{eq:Hamloc} accounts for the kinetic energy contribution of ions oscillating in place through the term proportional to the lattice momentum $\mathrm{P}$. It does not account for lattice distortion, or for energy transported in finite-wavevector phonon waves. As propagating phonons are analogous to propagating elastic waves we account for this additional energy by considering an additional contribution to the Hamiltonian of the lattice
\begin{equation}
	\hat{\mathcal{F}} = \bar{\mathcal{C}}_{ijkl} \hat{\mathrm{S}}_{ij} \hat{\mathrm{S}}_{kl}, \label{eq:sfe}
\end{equation}
where $\bar{\mathcal{C}}_{ijkl}$ is an effective elasticity tensor satisfying the symmetry conditions \cite{Gubbin2020b}
\begin{equation}
	\bar{\mathcal{C}}_{ijkl} = \bar{\mathcal{C}}_{jikl} = \bar{\mathcal{C}}_{klji}, \label{eq:sym}
\end{equation} 
and the $\hat{\mathrm{S}}_{ij}$ are scalar stresses operators defined as
\begin{equation}
	\hat{\mathrm{S}}_{ij} = \frac{1}{2} \left[ \frac{\mathrm{d \hat{X}}_i}{\mathrm{d} r_j} + \frac{\mathrm{d \hat{X}}_j}{\mathrm{d} r_i}\right].
\end{equation}
Note that this is a first order approximation to the elastic free energy, which will result in a quadratic phonon dispersion relation. This approximation is at the same level as that utilised in classical theories of polar nonlocality \cite{Gubbin2020}. The full nonlocal Hamiltonian is a sum of the two Hamiltonian densities
\begin{equation}
	\hat{\mathcal{H}} = \int \mathrm{d^3 r} \left[ \hat{\mathcal{H}}_0 + \hat{\mathcal{F}}\right]. \label{eq:Hamfull}
\end{equation}

\subsection{Equations of Motion}
The goal of this paper is to diagonalise Eq.\ref{eq:Hamfull}, writing it in terms of a series of bosonic ladder operators which describe the polaritonic eigenmodes of the system. To that end following previous approaches \cite{Gubbin2016b} we introduce the general polaritonic operator as a linear superposition of the free fields
\begin{equation}
	\hat{\mathcal{K}} = \int \mathrm{d^3 r} \left[ \boldsymbol{\alpha} \cdot \hat{\mathbf{D}}  + \boldsymbol{\beta} \cdot \hat{\mathbf{H}} +\boldsymbol{\gamma} \cdot \hat{\mathbf{X}}  + \boldsymbol{\zeta} \cdot \hat{\mathbf{Q}}\right], \label{eq:polaritondef}
\end{equation}
where Greek symbols are Hopfield fields describing the weighting of the fields comprising the eigenmode. If $\hat{\mathcal{K}}$ is an eigenmode of $\hat{\mathcal{H}}$ it satisfies the equation of motion
\begin{equation}
	\left[\hat{\mathcal{H}}, \hat{\mathcal{K}}\right] = \hbar \omega \hat{\mathcal{K}}, \label{eq:eom}
\end{equation}
where $\omega$ is the polariton frequency. The lengthy quantisation procedure, carried out in Appendix A, yields equations of motion for the Hopfield fields of the polariton
\begin{subequations}
	\begin{align}
		\omega \boldsymbol{\theta} &= i  \frac{\kappa  \boldsymbol{\zeta}}{\epsilon_0 \epsilon_{\infty}} - \frac{i  c^2}{\epsilon_{\infty}} \nabla \times \boldsymbol{\beta}, \label{eq:Nleom1} \\
		\omega \boldsymbol{\beta} &= i  \nabla \times \boldsymbol{\theta}, \label{eq:Nleom2}\\
		\omega \boldsymbol{\zeta} &= \frac{i}{\rho}  \boldsymbol{\gamma}, \\
		\omega \boldsymbol{\gamma} &= - i  \omega_{\mathrm{L}}^2 \rho \boldsymbol{\zeta} + \frac{i  c^2 \kappa}{\epsilon_{\infty}} \nabla \times \boldsymbol{\beta} + i \rho^{-1} \nabla\cdot \bar{\boldsymbol{\tau}}, \label{eq:Nleomn}
	\end{align}	
\end{subequations}
where $\bar{\boldsymbol{\tau}}$ is the stress tensor of the polar lattice, related to the effective elasticity tensor through
\begin{equation}
	\tau_{ij} = \frac{\bar{\mathcal{C}}_{ijkl} \rho}{4} \left( \frac{\partial \zeta_k}{\partial r_l}  + \frac{\partial \zeta_l}{\partial r_k}\right), \label{eq:elacons}
\end{equation}
and we defined the composite field
\begin{equation}
	\boldsymbol{\theta} = \boldsymbol{\alpha} + \frac{\left[\kappa  \boldsymbol{\zeta} \right]^{\mathrm{L}}}{\epsilon_0 \epsilon_{\infty}}.
\end{equation}
This field can be interpreted as the sum of the displacement field associated with the polariton's transverse photonic component, proportional to $\boldsymbol{\alpha}$ and the longitudinal electric field associated with the polariton's LO  phonon component, given by the latter term. The result is a new field $\boldsymbol{\theta}$ which acts as the full electric field of the polariton.\\
Note that in these equations of motion we can immediately identify the Hopfield fields as analogues of the classical fields. The magnetic field analogue is $\boldsymbol{\beta}$, while the electric field is $\boldsymbol{\theta}$. The other coefficients refer to the ionic position and momentum. It will be useful to the reader to keep this interpretation in mind through the rest of this manuscript.
%We can write the final equation
%\begin{equation}
%	\left(\omega_{\mathrm{T}}^2 - \omega^2 \right) \boldsymbol{\zeta} +  \nabla\cdot \bar{\boldsymbol{\tau}} =  i \kappa \omega \boldsymbol{\theta}, \label{eq:finaleom}
%\end{equation}
%which is the nonlocal equation of motion for the Hopfield fields. 
We now show that the equations of motion lead to the dispersion relations predicted by classical approaches \cite{Gubbin2020}. For simplicity we focus on isotropic crystal structures, for which the components of $\bar{\mathcal{C}}_{ijkl}$ are given by
\begin{align}
	\lambda_1 = \bar{\mathcal{C}}_{1111} &= \bar{\mathcal{C}}_{2222} = \bar{\mathcal{C}}_{3333}, \\
	\lambda_2 = \bar{\mathcal{C}}_{1122} &= \bar{\mathcal{C}}_{1133} = \bar{\mathcal{C}}_{2233} = \dots,\\
	\lambda_3 = \bar{\mathcal{C}}_{1212} &= \bar{\mathcal{C}}_{2112} = \bar{\mathcal{C}}_{1313} = \dots,
\end{align}
where the dots indicate other permutations of indices through Eq.~\ref{eq:sym}. These components relate to the longitudinal and transverse phonon velocities in the quadratic dispersion limit $\beta_{\mathrm{L}}, \; \beta_{\mathrm{T}}$ introduced phenomenologically in other works \cite{Gubbin2020, Gubbin2021} through
\begin{align}
	\lambda_1 &= \rho \beta_{\mathrm{L}}^2,\\
	\lambda_2 &= \rho \left(\beta_{\mathrm{L}}^2 - 2 \beta_{\mathrm{T}}^2\right),\\
	\lambda_3 &= \rho \beta_{\mathrm{T}}^2.
\end{align}
Finally through Eq.~\ref{eq:elacons} the components of the effective stress tensor can be derived
\begin{align}
	\tau_{11} &=  \rho \left[\beta_{\mathrm{L}}^2 \mathcal{S}_{11} + \left(\beta_{\mathrm{L}}^2 - 2 \beta_{\mathrm{T}}^2 \right) \left( \mathcal{S}_{22} +  \mathcal{S}_{33}\right) \right],\\
	\tau_{22} &=  \rho \left[\beta_{\mathrm{L}}^2  \mathcal{S}_{22} + \left(\beta_{\mathrm{L}}^2 - 2 \beta_{\mathrm{T}}^2 \right) \left( \mathcal{S}_{11} +  \mathcal{S}_{33}\right) \right],\\
	\tau_{33} &=  \rho \left[\beta_{\mathrm{L}}^2  \mathcal{S}_{33} + \left(\beta_{\mathrm{L}}^2 - 2 \beta_{\mathrm{T}}^2 \right) \left( \mathcal{S}_{11} +  \mathcal{S}_{22}\right) \right],\\
	\tau_{12} &= \tau_{21} = \rho \beta_{\mathrm{T}}^2  \mathcal{S}_{12},\\
	\tau_{13} &= \tau_{31} = \rho \beta_{\mathrm{T}}^2  \mathcal{S}_{13},\\
	\tau_{23} &= \tau_{32} = \rho \beta_{\mathrm{T}}^2  \mathcal{S}_{23},
\end{align}
where the $ \mathcal{S}_{ij}$ are scalar stresses of the Hopfield field $\zeta$
\begin{equation}
	\mathcal{S}_{ij} = \frac{\rho}{2} \left[ \frac{\mathrm{d} \zeta_i}{\mathrm{d} r_j} + \frac{\mathrm{d} \zeta_j}{\mathrm{d} r_i}\right].
\end{equation}

One class of solutions of  Eq.~\ref{eq:Nleom1}-\ref{eq:Nleomn} are transverse fields, for which the Hopfield field for the classical electric field satisfies $\nabla \cdot \boldsymbol{\theta} = 0$. In this case we can solve the equations of motion directly, recovering the nonlocal wave equation for transverse fields
\begin{align}
	\nabla\times\nabla\times\boldsymbol{\theta} &= \frac{\omega^2 \epsilon_{\mathrm{T}}\left(\omega, k\right)}{c^2} \boldsymbol{\theta}. \label{eq:traweq}
\end{align}
where $\epsilon_{\mathrm{T}}$ is the transverse dielectric function of the lattice, given by
\begin{equation}
	\epsilon_{\mathrm{T}} \left(\omega, k\right) = \epsilon_{\infty} \frac{\omega_{\mathrm{L}}^2 - \omega^2 - \beta_{\mathrm{T}}^2 k^2}{\omega_{\mathrm{T}}^2 - \omega^2 - \beta_{\mathrm{T}}^2 k^2}, \label{eq:epstra}
\end{equation}
and $\beta_{\mathrm{T}}$ can be interpreted as the transverse optical phonon velocity in the quadratic limit. \\
The equations of motion also support longitudinal solutions for which the classical electric Hopfield field satisfies $\nabla \times \boldsymbol{\theta} = 0$. This results in dispersion relation
\begin{equation}
	\epsilon_{\mathrm{L}} \left(\omega, k\right) = 0,	
\end{equation}
where $\epsilon_{\mathrm{L}}$ is the longitudinal dielectric function of the lattice, given by
\begin{equation}
	\epsilon_{\mathrm{L}} \left(\omega, k\right) = \epsilon_{\infty} \frac{\omega_{\mathrm{L}}^2 - \omega^2 - \beta_{\mathrm{L}}^2 k^2}{\omega_{\mathrm{T}}^2 - \omega^2 - \beta_{\mathrm{L}}^2 k^2},
\end{equation}
where $\beta_{\mathrm{L}}$ is the transverse optical phonon velocity. These transverse and longitudinal dispersion relations are exactly those predicted from the classical theory, validating the approach taken in this section. The LTP excitations which are the focus of this work are hybrid modes with both transverse and longitudinal components, mixed by the boundary conditions on the Hopfield fields $\boldsymbol{\alpha, \beta, \gamma, \zeta}$ which must enforce the continuity of the quantum fields as predicted in standard theories of polar nonlocality \cite{Gubbin2020}. These boundary conditions are derived in Sec.~\ref{sec:Bcs}

\subsection{Field Construction}
We can solve the equations of motion Eq.~\ref{eq:Nleom1}-\ref{eq:Nleomn} in terms of a set of eigenmodes
\begin{equation}
    \lvert \boldsymbol{\Psi}_n \rangle = \lvert \boldsymbol{\alpha}_n, \boldsymbol{\beta}_n, \boldsymbol{\gamma}_n, \boldsymbol{\zeta}_n \rangle,
\end{equation}
where the index $n$ can be either discrete, or continuous, depending on the mode and system under study with associated eigenfrequencies $\omega_n$ and wavevectors $k_n$. Then an arbitrary field operator $\hat{\mathbf{Y}}$ can be written as a sum over the polaritonic operators as
\begin{equation}
	\hat{\mathbf{Y}} = \sum_n \left[ \mathbf{f}_n^{\mathrm{Y}} \hat{\mathcal{K}}_n +  \bar{\mathbf{f}}_n^{\mathrm{Y}} \hat{\mathcal{K}}_n^{\dag}\right], 
\end{equation}
in which the expansion coefficients are given by
\begin{equation}
	\mathbf{f}_n^{\mathrm{Y}} = \left[\hat{\mathbf{Y}}, \hat{\mathcal{K}}_{n}^{\dag} \right].
\end{equation}
The electric displacement field, magnetic field, ionic displacement, electric field operator $\hat{\mathbf{E}}$ and polarisation field operator $\hat{\mathbf{P}}$ can be written
\begin{subequations}
\begin{align}
    \hat{\mathbf{D}} &= - \sum_n \hbar \omega_n \epsilon_0   \epsilon_\mathrm{T} \left(\omega_n,  k_n\right) \left[ \bar{\boldsymbol{\theta}}_n \hat{\mathcal{K}}_n +  \boldsymbol{\theta}_n \hat{\mathcal{K}}_n^{\dag}\right], \label{eq:dfield}\\
     \hat{\mathbf{H}} &= - \sum_n \frac{\hbar \omega_n}{\mu_0} \left[\bar{\boldsymbol{\beta}}_n \hat{\mathcal{K}}_n +   \boldsymbol{\beta}_n \hat{\mathcal{K}}_n^{\dag}\right], \label{eq:hfield}\\
    \hat{\mathbf{X}} &= i \hbar  \sum_n \left[  \bar{\boldsymbol{\zeta}}_n \hat{\mathcal{K}}_n -   \boldsymbol{\zeta}_n \hat{\mathcal{K}}_n^{\dag}\right], \label{eq:xfield}\\
    \hat{\mathbf{P}} &= i \hbar \kappa \sum_n \left[  \bar{\boldsymbol{\zeta}}_n \hat{\mathcal{K}}_n -   \boldsymbol{\zeta}_n \hat{\mathcal{K}}_n^{\dag}\right] \nonumber \\
    &= i \hbar \epsilon_0 \sum_n \omega_n \left[ \epsilon_{\infty} - \epsilon_{\mathrm{T}}\left(\omega_n, k\right)\right]\left[  \bar{\boldsymbol{\theta}}_n \hat{\mathcal{K}}_n -   \boldsymbol{\theta}_n \hat{\mathcal{K}}_n^{\dag}\right], \label{eq:pfield}\\
    \hat{\mathbf{E}} &= \frac{1}{\epsilon_0 \epsilon_{\infty}} \left[\hat{\mathbf{D}} - \hat{\mathbf{P}}\right] =  - \sum_n \hbar \omega_n \left[ \bar{\boldsymbol{\theta}}_n \hat{\mathcal{K}}_n +  \boldsymbol{\theta}_n \hat{\mathcal{K}}_n^{\dag}\right], \label{eq:efield}
\end{align}
\end{subequations}
where in the final relation we used the relation between the material polarisation density operator $\hat{\mathbf{P}} = \kappa \hat{\mathbf{X}}$ and the following relation derived from Eq.~\ref{eq:Nleom1}-\ref{eq:Nleom2} by elimination of $\boldsymbol{\beta}$
\begin{equation}
    i \kappa \boldsymbol{\zeta} = \epsilon_0 \omega \left[\epsilon_{\infty} - \epsilon_{\mathrm{T}} \left(\omega, k\right) \right]\boldsymbol{\theta}.
\end{equation}
Again note the physical significance of the polariton Hopfield fields. The displacement field is proportional to the electric Hopfield field $\boldsymbol{\theta}$ multiplied by the system's transverse dielectric function, and the magnetic field operator is proportional to the magnetic Hopfield field $\boldsymbol{\beta}$.

\subsection{Boundary Conditions}
\label{sec:Bcs}
We have derived the equations of motion for the nonlocal polariton system Eq.\ref{eq:Nleom1}-\ref{eq:Nleomn}. This equation set describes the relation between the Hopfield fields in a homogeneous system. To solve a physical problem we also need to derive boundary conditions, describing how polaritonic Hopfield fields behave at material interfaces. As has been discussed at length in previous works the standard boundary conditions utilised in electromagnetic theory are insufficient to describe nonlocal systems due to the introduction of additional fields describing the lattice distortion \cite{Gubbin2020}. The appropriate new boundary conditions can be derived starting from the classical Poynting relation which links the rate of change of electromagnetic energy density in a volume $\Omega$ with the energy flux passing through enclosing surface $\mathrm{d}\Omega$ with outgoing surface normal unit vector $\overrightarrow{\mathbf{n}}$, given by
\begin{multline}
    \int_{\mathrm{d} \Omega} \left[\hat{\mathbf{E}} \times \hat{\mathbf{H}} \right] \cdot \overrightarrow{\mathbf{n}} \mathrm{d S} \\
    = - \int_{\Omega} \left[ \epsilon_0 \epsilon_{\infty} \hat{\mathbf{E}} \cdot \dot{\hat{\mathbf{E}}} + \kappa \hat{\mathbf{E}} \cdot \dot{\hat{\mathbf{X}}} + \mu_0 \hat{\mathbf{H}} \cdot \dot{\hat{\mathbf{H}}}\right] \mathrm{d V} , \label{eq:poynting}
\end{multline}
where we utilised the standard constitutive relation for a nonlocal dielectric to eliminate the displacement field \cite{Gubbin2020}. In this equation terms proportional to a field multiplied by it's time derivative describe the rate of change of energy densities in $\Omega$. The term on the right proportional to $\kappa$ can be expressed as the sum of such a density and a transport term utilising the following relation, derived by elimination of $\boldsymbol{\gamma, \beta}$ from Eq.~\ref{eq:Nleomn}
\begin{equation}
    i \omega \kappa \boldsymbol{\theta} =  \rho \left[\omega_{\mathrm{T}}^2 -\omega^2 \right]\boldsymbol{\zeta}  + \rho \nabla\cdot \bar{\boldsymbol{\tau}},
\end{equation}
to obtain
\begin{align}
    \kappa \hat{\mathbf{E}} \cdot \dot{\hat{\mathbf{X}}} &= - \sum_n \omega_n \hbar^2 \kappa \boldsymbol{\theta}_n \cdot \dot{\boldsymbol{\zeta}}_n \;\; \hat{\mathcal{K}}_{n}^{\dag} \hat{\mathcal{K}}_{n}^{\dag} + \dots \nonumber \\
    &= \sum_n i\hbar  \left[\rho \left[\omega_{\mathrm{T}}^2 -\omega_n^2 \right]\boldsymbol{\zeta}_n  + \rho \nabla\cdot \bar{\boldsymbol{\tau}}_n \right]  \cdot \dot{\boldsymbol{\zeta}}_n \;\; \hat{\mathcal{K}}_{n}^{\dag} \hat{\mathcal{K}}_{n}^{\dag} + \dots.
\end{align}
The dots represent terms proportional to the remaining combinations of operators ($\hat{\mathcal{K}}_{n}^{\dag} \hat{\mathcal{K}}_{n}, \hat{\mathcal{K}}_{n} \hat{\mathcal{K}}_{n}^{\dag}, \hat{\mathcal{K}}_{n} \hat{\mathcal{K}}_{n}$). For Eq.~\ref{eq:poynting} to be valid terms proportional to each permutation of operators on each side must balance. Rewriting the volume integral of the second term we find
\begin{align}
    &\int_{\Omega} \mathrm{d V} \rho \nabla \cdot \bar{\boldsymbol{\tau}}_{n,i}  \cdot \dot{\zeta}_{n,i}\nonumber \\
    &\quad \quad = \int_{\Omega} \mathrm{d V} \rho \left[\nabla \cdot \left(\dot{\zeta}_{n,i} \bar{\boldsymbol{\tau}}_{n,i}\right) - \bar{\boldsymbol{\tau}}_i \nabla \dot{\zeta}_{n, i} \right] \nonumber \\
    & \quad \quad = \int_{\mathrm{d} \Omega} \mathrm{d S} \rho \left[\dot{\zeta}_{n,i} \bar{\boldsymbol{\tau}}_{n,i}\right] \cdot \overrightarrow{\mathbf{n}} - \int_{\Omega} \mathrm{d V} \rho \bar{\boldsymbol{\tau}}_{n,i} \cdot \nabla \dot{\zeta}_{n, i},
\end{align}
where index $i$ refers to the cartesian components of $\boldsymbol{\zeta}_n$. The remaining volume integral describes energy stored in elastic distortions of the lattice, while the surface integral describes energy transported in finite wavevector phonon modes. Bringing the surface integral over to the left write a composite Poynting-like vector for the Hopfield fields 
\begin{equation}
    \boldsymbol{\sigma}_n = \frac{\hbar^2}{\mu_0} \omega_n^2 \left(\boldsymbol{\theta}_n \times \boldsymbol{\beta}_n\right) + \rho \bar{\boldsymbol{\tau}}_{n}\dot{\boldsymbol{\zeta}}_{n},
\end{equation}
where we collected the components proportional to $\hat{\mathcal{K}}_n^{\dag}\hat{\mathcal{K}}_n^{\dag}$ and expanded the electromagnetic Poynting vector utilising Eq.~\ref{eq:hfield}, \ref{eq:efield} as
\begin{equation}
    \left[\hat{\mathbf{E}} \times \hat{\mathbf{H}}\right] \cdot \overrightarrow{\mathbf{n}} = \frac{\hbar^2}{\mu_0} \sum_n \omega_n^2 \left(\boldsymbol{\theta}_n \times \boldsymbol{\beta}_n\right) \cdot \overrightarrow{\mathbf{n}} \;\; \hat{\mathcal{K}}_{n}^{\dag} \hat{\mathcal{K}}_{n}^{\dag} + \dots,
\end{equation}
where as for the right hand side we collect terms proportional to $\hat{\mathcal{K}}_{n}^{\dag} \hat{\mathcal{K}}_{n}^{\dag}$. Continuity of $\boldsymbol{\sigma}_n$ requires that tangential components of $\boldsymbol{\theta}_n, \boldsymbol{\sigma}_n$ are continuous across material interfaces. As can be seen from inspection of Eq.~\ref{eq:dfield}-\ref{eq:hfield} these are the standard Maxwell boundary conditions on the electromagnetic fields $\mathbf{E, H}$. The second term fixes the additional boundary conditions for the nonlocal problem, requiring that $\boldsymbol{\zeta}_n$ and the normal component of $\bar{\boldsymbol{\tau}}_n$ are continuous. The former from Eq.~\ref{eq:xfield} enforces continuity of the lattice ionic displacement, while the latter constrains the parallel and perpendicular components of the stress at the interface, both in agreement with classical theories of polar nonlocality \cite{Gubbin2020}. Physically the additional nonlocal boundary conditions are required to describe the additional phonon fields in the dielectric, and to weight them with respect to the photonic fields. They determine the degree of nonlocal mixing.

\subsection{LTP Quantisation}
Now we have all the tools necessary to find the Hopfield fields for a given inhomogeneous system. There is still one step remaining, as for operators $\hat{\mathcal{K}}_n$ to describe fundamental bosonic excitations of the nonlocal system they also need to be quantised as bosonic fields \cite{Gubbin2016b}
\begin{equation}
	\left[\hat{\mathcal{K}}_m, \hat{\mathcal{K}}_n^{\dag}\right] = \delta_{mn} \mathrm{sgn}\left(\omega_n\right),
\end{equation}
where the sign function yields $\mathrm{sgn}\left(\omega_n\right) = \omega_n / \lvert \omega_n\rvert$. Utilising the definition of the polaritonic operators in terms of the quantum fields, we derive the result
\begin{align}
	i \hbar \int \mathrm{d^3 r} \biggr[&\frac{1}{\mu_0} \boldsymbol{\alpha}_m \cdot \left(\nabla \times \bar{\boldsymbol{\beta}}_n\right) -  \frac{1}{\mu_0} \left(\nabla \times \boldsymbol{\beta}_m\right) \cdot \bar{\boldsymbol{\alpha}}_n \nonumber \\
	&  + \frac{1}{\rho}\left(\boldsymbol{\gamma}_m \cdot \bar{\boldsymbol{\zeta}}_n - \boldsymbol{\zeta}_m \cdot \bar{\boldsymbol{\gamma}}_n \right)\biggr] = \delta_{mn} \mathrm{sgn}\left(\omega_n\right).\label{eq:polquant}
\end{align}

To proceed it is useful to introduce a further layer of abstraction. Although an LTP is comprised of distinct longitudinal and transverse parts this is not captured by the Hopfield fields $\lvert \boldsymbol{\Psi}_n \rangle = \lvert \boldsymbol{\alpha}_n, \boldsymbol{\beta}_n, \boldsymbol{\gamma}_n, \boldsymbol{\zeta}_n \rangle$ which describe the fully-coupled fields of the polariton. We can obtain a more physically meaningful result by segregating the fields into longitudinal and transverse components whose electric Hopfield fields satisfy $\nabla \cdot \boldsymbol{\theta}_{n}^{\mathrm{T}} = 0$ and $\nabla \times \boldsymbol{\theta}_{n}^{\mathrm{L}} = 0$ respectively. We can then quantise these components separately. For transverse fields  utilising the transverse dielectric function defined in Eq.~\ref{eq:epstra}, we find
\begin{multline}
	 \text{sgn}\left(\omega_n\right) \\
	 = \hbar \omega_n \epsilon_0 \int \mathrm{d^3 r} \lvert \theta_n^{\mathrm{T}} \rvert^2 \left[ \epsilon_{\mathrm{T}}\left(\omega, k\right) + \frac{\partial\left[\omega  \epsilon_{\mathrm{T}}\left(\omega, k\right)\right]}{\partial \omega} \right]_{\omega = \omega_n}.  \label{eq:traquant}
\end{multline}
Note that utilising 
\begin{align}
	\omega^2 \lvert \beta^{\mathrm{T}} \rvert^2 &= \left[ \nabla \times \boldsymbol{\theta}^{\mathrm{T}} \right] \left[\nabla \times \boldsymbol{\theta}^{\mathrm{T}*}\right] = \frac{\omega^2 \epsilon_{\mathrm{T}}\left(\omega, k\right)}{c^2} \lvert \theta^{\mathrm{T}} \rvert^2,
\end{align}
we can re-write this as in terms of both the electric and magnetic Hopfield fields as
\begin{multline}
	 \mathrm{sgn}\left(\omega_n\right) \\= 4 \hbar \omega_n \int \mathrm{d^3 r} \biggr[\frac{\epsilon_0}{4}  \frac{\partial\left[\omega  \epsilon_{\mathrm{T}}\left(\omega, k\right)\right]}{\partial \omega}    \lvert \theta_n^{\mathrm{T}} \rvert^2 + \frac{1}{4 \mu_0} \lvert \beta_n^{\mathrm{T}} \rvert^2 \biggr]_{\omega = \omega_n}, \label{eq:traquantb}
\end{multline}
which in the local limit $k\to 0$ is the transverse field energy density in a dispersive dielectric \cite{Ruppin2002}.\\
For longitudinal fields the Hopfield field $\boldsymbol{\beta}_n^{\mathrm{T}}$, which gives the magnetic field contribution to the polariton goes to zero so we can write
\begin{align}
	2 \hbar \omega_n \epsilon_0 \int \mathrm{d^3 r} \frac{\epsilon_{\infty} \omega_n^2 \left[\omega_{\mathrm{L}}^2 - \omega_{\mathrm{T}}^2\right]}{\left[ \omega_n^2  -  \omega_{\mathrm{T}}^2  + \beta_{\mathrm{L}}^2 k_n^2\right]^2} \lvert \theta_n^{\mathrm{L}}\rvert^2 = \mathrm{sgn}\left(\omega_n\right), \label{eq:LOq1}
\end{align}
where we assumed an isotropic lattice. Note that for a pure longitudinal excitation, whose frequency satisfies $\epsilon_{\mathrm{L}}\left(k, \omega\right) = 0$ this can be simplified to
\begin{equation}
	4  \hbar \omega_n \epsilon_0 \int \mathrm{d^3 r}\frac{ \epsilon_{\rho} \omega_n^2 }{2 \omega_\mathrm{L}^2  } \lvert \theta_n^{\mathrm{L}}\rvert^2 =  \mathrm{sgn}\left(\omega_n\right),
\end{equation}
where we defined the Fr{\"o}hlich coupling constant utilising the Lydanne-Sachs-Teller relation in Eq.~\ref{eq:LST}
\begin{equation}
	\epsilon_{\rho} = \left[\frac{1}{\epsilon_{\infty}} -  \frac{1}{\epsilon_{\mathrm{st}}}\right]^{-1}.
\end{equation}
It is important to make clear that in these quantisation relations the frequencies are those of the hybrid LTP, determined by application of the full boundary conditions, not the individual longitudinal or transverse fields \cite{Gubbin2021b}. Finally having quantised these excitations we introduce additional polariton-like Hopfield fields describing the longitudinal (transverse) composition of the $n$th LTP mode $\mathrm{L}_{i, n} \; (\mathrm{T}_{j, n})$ which satisfy the normalisation
\begin{equation}
    \sum_{i} \lvert \mathrm{L}_{i, n} \rvert^2 + \sum_j \lvert \mathrm{T}_{j, n} \rvert^2 = 1,
\end{equation}
where the index $i$ rules over all the longitudinal components and $j$ over all transverse components of the polariton. Then if we expand the index of the Hopfield fields to consider both the LTP branch $n$ and the contributing mode $ij$ as $\boldsymbol{\alpha}_{i, n}^{\mathrm{L}}, \boldsymbol{\alpha}_{j, n}^{\mathrm{T}}$ satisfy Eq.~\ref{eq:polquant} the full polariton field given by
\begin{equation}
    \boldsymbol{\alpha}_{m} = \sum_{i}\mathrm{L}_{i, m} \boldsymbol{\alpha}_{i, m}^{\mathrm{L}} + \sum_j \mathrm{T}_{j, m} \boldsymbol{\alpha}_{j, m}^{\mathrm{T}},
\end{equation}
is also quantised.\\

\section{Purcell Enhanced Emission Near an Epsilon-Near-Zero Waveguide}
\begin{figure}
    \includegraphics[width=0.4\textwidth]{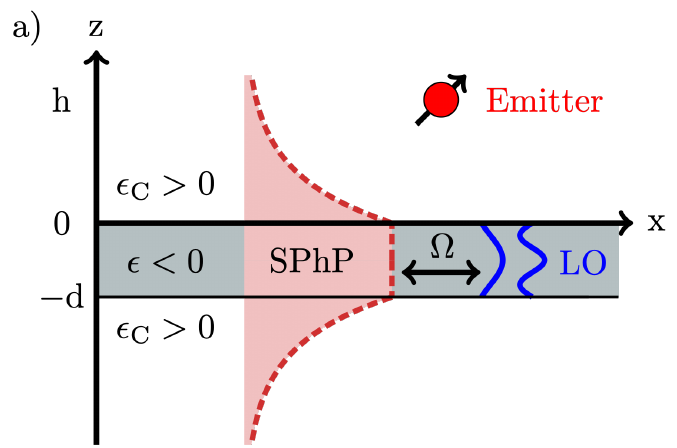}	
	\includegraphics[width=0.5\textwidth]{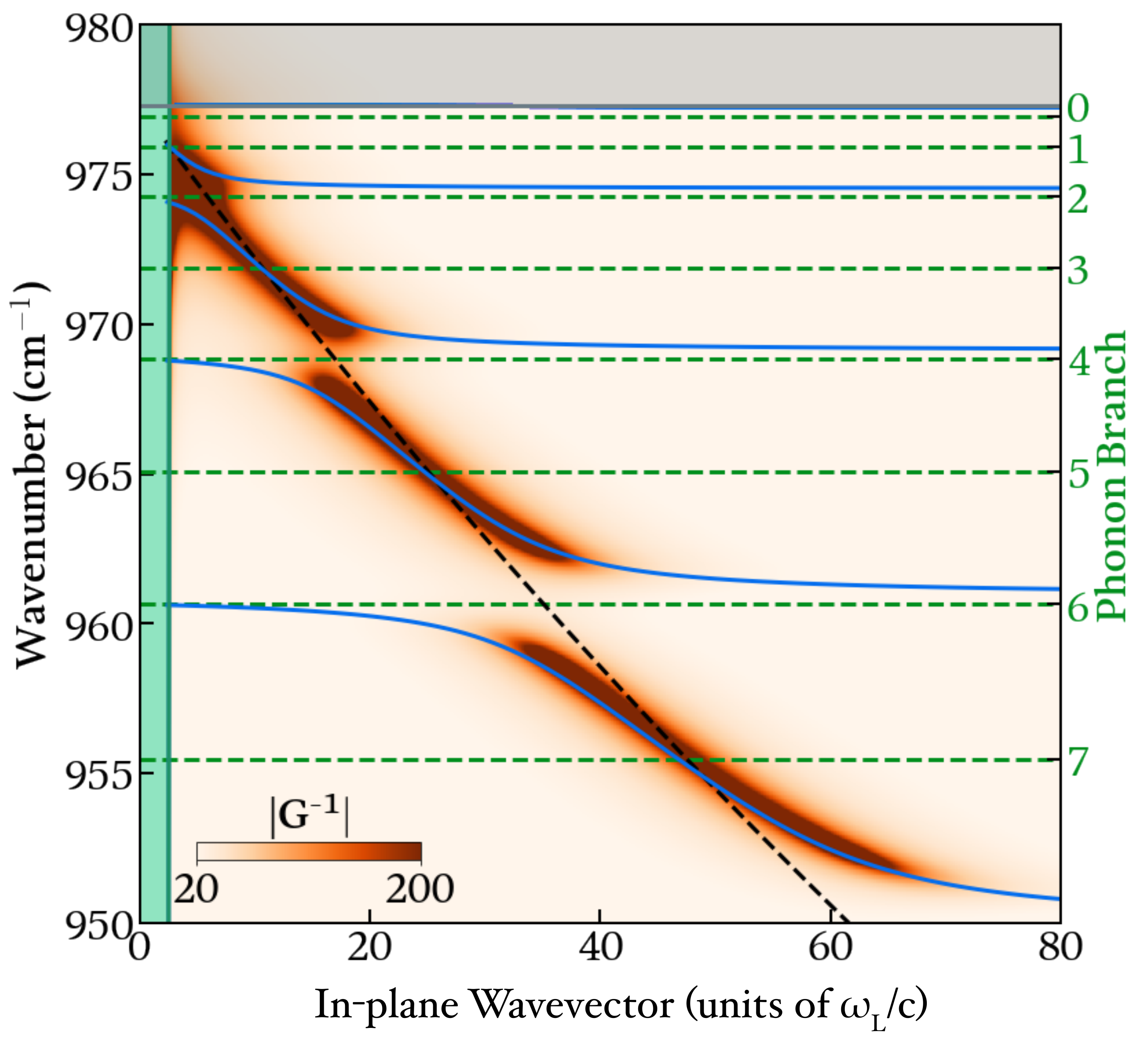}	
	\caption{\label{fig:Fig1} 
	a) A sketch of the thin film-emitter system. A negative dielectric film with transverse dielectric function $\epsilon_2 < 0$ sits between $-d<z<0$. Symmetric positive dielectric cladding with dielectric constant $\epsilon_1$ occupies $z < - d$ and $z > 0$. The film supports an SPhP excitation (red) and a discrete spectrum of localised LO phonon modes (blue) which are hybridised through the film boundaries with coupling frequency $\Omega$. We consider an emitter embedded in the upper, positive dielectric, halfspace, separated from the film by a height $h$. 
	b) The inverse of the polariton dispersion $G$ as defined in Appendix D for a 10nm 3C-SiC film between two high-index ($n_{\mathrm{C}} = 2.6$) cladding regions. Dark regions illustrate the modal frequencies. Horizontal dashed green lines indicate the LO phonon branches, black dashed line indicates the epsilon near zero dispersion. Polariton frequencies are shown by the blue solid lines.}
\end{figure}
Now we apply our theory to a physical system. We focus on the system of an isotropic polar dielectric film of thickness $d$ sandwiched between positive dielectric cladding layers. For simplicity we consider the case where only nonlocal longitudinal excitations are present in the system, a good approximation for systems driven near the LO phonon frequency where the TO phonon dispersion does not extend. The system under study is shown in Fig.~\ref{fig:Fig1}a. The film, considered to occupy $-d < z < 0$ has dispersive dielectric function $\epsilon$, while the cladding regions defined as $z < -d, z > 0$ have frequency independent dielectric constant $\epsilon_{\mathrm{C}}$.\\
In the thick film limit this kind of symmetric planar waveguide supports two spectrally degenerate SPhP modes, with the standard dispersion for the evanescent waves on a planar bilayer. When the film thickness $d$ is less than the skin depth of the modes the SPhPs on each interface hybridise into symmetric and antisymmetric superpositions, lifting the spectral degeneracy. As a result of repulsion between like charges on each side of the film the symmetric mode blue shifts toward the LO phonon frequency which marks the upper edge of the Reststrahlen region, in which evanescent modes are supported. In the thin film limit this mode sits close to the LO phonon frequency where the polar film's local dielectric function goes to zero. For this reason it is termed an epsilon-near-zero excitation. These modes are especially interesting in nanophotonics as they allow for strong enhancement of the out-of-plane field in the nanolayer with applications in sensing and mid-infrared nano-waveguiding \cite{Passler2018, Campione2015}. These systems are especially interesting in optical nonlocality, where classical studies ahve demonstrated hybridisation between them and localised longitudinal phonons in the nanolayer \cite{Gubbin2020, Gubbin2021}.\\

\subsection{Modes of the System}
We construct the eigenmode of the epsilon-near-zero mode to satisfy the Maxwell boundary condition on the magnetic Hopfield field $\boldsymbol{\beta}$ as the LO field has no associated $\boldsymbol{\beta}$ field. The eigenmode is a sum of SPhPs localised at each film/cladding interface. We denote the wavevector in the plane of the waveguide as $\mathbf{k} = k_x \overrightarrow{\mathbf{x}} + k_y \overrightarrow{\mathbf{y}}$, and for brevity write the 3D wavevector in form $\left[\mathbf{k}, k_z\right]$ where $k_z$ is the out-of-plane component and the position $\left[\mathbf{r}, z\right]$. The electric Hopfield field $\boldsymbol{\theta}$ associated with the SPhP localised at $z = -d$ is given by
\begin{equation}
    \boldsymbol{\theta}_{u, \mathbf{k}} = \begin{cases}
    - \frac{\alpha_\mathrm{C} \epsilon}{\alpha \epsilon_{\mathrm{C}}} \left[\frac{\mathbf{k}}{\lvert \mathbf{k} \rvert}, - \frac{i \lvert \mathbf{k} \rvert}{\alpha_\mathrm{C}} \right]^{\mathrm{T}}  e^{\alpha_\mathrm{C} \left(z + d\right)} e^{i \mathbf{k} \cdot \mathbf{r}} & z < -d,\\
	\left[\frac{\mathbf{k}}{\lvert \mathbf{k} \rvert},  \frac{i \lvert\mathbf{k}\rvert}{\alpha}\right]^{\mathrm{T}} e^{-\alpha \left(z + d\right)} e^{i \mathbf{k} \cdot \mathbf{r}}  & -d < z < 0,\\
	\frac{\alpha_\mathrm{C} \epsilon}{\alpha \epsilon_\mathrm{C}} \left[\frac{\mathbf{k}}{\lvert \mathbf{k} \rvert},  \frac{i \lvert \mathbf{k}\rvert}{\alpha_\mathrm{C}}\right]^{\mathrm{T}} e^{-\alpha_\mathrm{C} z - \alpha d} e^{i \mathbf{k} \cdot \mathbf{r}}  & z > 0,
	\end{cases}
\end{equation}
while that located at $z = 0$ has 
\begin{equation}
    \boldsymbol{\theta}_{l, \mathbf{k}} = \begin{cases}
    -  \frac{\alpha_\mathrm{C} \epsilon}{\alpha \epsilon_\mathrm{C}} \left[\frac{\mathbf{k}}{\lvert \mathbf{k} \rvert}, - \frac{i \lvert \mathbf{k} \rvert}{\alpha_\mathrm{C}} \right]^{\mathrm{T}} e^{\alpha_\mathrm{C} \left(z + d\right) - \alpha d} e^{i \mathbf{k} \cdot \mathbf{r}} & z < -d,\\
	-  \left[\frac{\mathbf{k}}{\lvert \mathbf{k} \rvert},  -\frac{i \lvert \mathbf{k} \rvert}{\alpha}\right]^{\mathrm{T}} e^{\alpha z } e^{i \mathbf{k} \cdot \mathbf{r}} & -d < z < 0,\\
	\frac{\alpha_\mathrm{C} \epsilon}{\alpha \epsilon_\mathrm{C}} \left[\frac{\mathbf{k}}{\lvert \mathbf{k} \rvert},  \frac{i \lvert \mathbf{k} \rvert}{\alpha_\mathrm{C}}\right]^{\mathrm{T}} e^{-\alpha_\mathrm{C} z } e^{i \mathbf{k} \cdot \mathbf{r}} & z > 0,
	\end{cases}
\end{equation}
In the above equations $\alpha \; (\alpha_\mathrm{C})$ is the out-of-plane wavevector in the film (cladding) 
\begin{align}
    \alpha &= \sqrt{\lvert \mathbf{k} \rvert^2 - \epsilon \frac{\omega^2}{c^2}},\\
    \alpha_{\mathrm{C}} &= \sqrt{\lvert \mathbf{k} \rvert^2 - \epsilon_{\mathrm{C}} \frac{\omega^2}{c^2}}.
\end{align}
The total symmetric field electric field coefficient in the waveguide is the linear superposition $\boldsymbol{\theta}^{\mathrm{T}} = \boldsymbol{\theta}_{u, \mathbf{k}} + \boldsymbol{\theta}_{l, \mathbf{k}}$. This collective excitation is quantised in Appendix B, and can be written in form
\begin{equation}
    \boldsymbol{\theta}_{\mathbf{k}}^{\mathrm{T}} = \mathrm{A}_{\mathbf{k}}^{\mathrm{T}} \mathbf{u}_{\mathbf{k}}\left(z\right) e^{i \mathbf{k} \cdot \mathbf{r}},
\end{equation}
where $\mathrm{A}_{\mathbf{k}}^{\mathrm{T}}$ is the quantisation factor, calculated by independent quantisation of the transverse field, and $\mathbf{u}_{\mathbf{k}}\left(z\right)$ are unit vectors describing the out-of-plane mode functions of the coupled excitations.\\

As the cladding layers are phonon-inactive the thin film also acts as a closed cavity for LO phonons, supporting modes with frequencies
\begin{equation}
    \omega_{\mathbf{k}, n}^{\mathrm{L}} = \sqrt{\omega_{\mathrm{L}}^2 - \beta_{\mathrm{L}}^2 \left(\lvert \mathbf{k} \rvert^2 + \xi_n^2\right)},
\end{equation}
where $n$ is an integer defining the mode number, $\xi_n = n \pi / d$ is the quantised out-of-plane wavevector of the phonon and $\mathbf{k}$ is the wavevector in the plane of the film. If we consider only symmetric modes in the waveguide $n \in \text{odd}$ the electric Hopfield field $\boldsymbol{\theta}$ of the phonons can be written
\begin{equation}
    \boldsymbol{\theta}_{\mathbf{k}, n}^{\mathrm{L}} = \mathrm{B}_{ \mathbf{k}, n}^{\mathrm{L}}\nabla \left\{e^{i \mathbf{k} \cdot \mathbf{r}} \sin \left[\xi_n (z + d / 2)\right]\right\},
\end{equation}
where $\mathrm{B}_{\mathbf{k}, n}^{\mathrm{L}}$ is a constant determined from the quantisation Eq.~\ref{eq:LOq1} derived in Appendix C and given by
\begin{align}
    \mathrm{B}_{\mathbf{k}, n}^{\mathrm{L}}  = \sqrt{\frac{\omega_{\mathrm{L}}^2}{\hbar \omega_{\mathbf{k}, n}^3 \epsilon_0 \epsilon_{\rho} \mathrm{S} \mathrm{L}_{\mathbf{k}, n}^{\mathrm{L}}}},
\end{align}
where $\mathrm{S}$ is the quantisation surface and $\mathrm{L}_{\mathbf{k}, n}^{\mathrm{L}}$ is a quantisation length for the mode also defined in Appendix C which accounts for the frequency shift of the mode from zone-centre.\\

Both sets of excitations are hybridised through the Maxwell boundary condition on the tangential component of $\boldsymbol{\theta}$ and the additional boundary conditions discussed in Section I-C. In keeping with classical studies of polar nonlocality we utilise the additional boundary condition on the normal component of $\boldsymbol{\zeta}$, which can be recast as a constraint on the normal component of $\epsilon_{\infty} \boldsymbol{\theta}$. Application of these boundary conditions yields dispersion relation
\begin{equation}
	\tanh\left(\frac{\alpha d}{2}\right) + \frac{\alpha_\mathrm{C} \epsilon}{\alpha \epsilon_\mathrm{C}}   = - \tanh\left( \frac{\xi_n d}{2}\right) \left(\frac{\epsilon}{\epsilon_{\infty}} - 1\right) \frac{\lvert \mathbf{k} \rvert^2}{\xi_n \alpha}, \label{eq:symdisp}
\end{equation}
which can be manipulated into form \cite{Gubbin2021}
\begin{equation}
	1 = \frac{\omega_{\mathrm{L}}^2 - \omega_{\mathbf{k}}^{\mathrm{T}\;2}}{\omega_\mathbf{k}^{\mathrm{T}\;2} - \omega^2} \sum_n \frac{8 \beta_{\mathrm{L}}^2 / d^2}{\omega_{\mathbf{k}, n}^{\mathrm{L}\;2} - \omega^2}.
\end{equation}
by expanding the hyperbolic functions into Mittag-Leffler series and exploiting the analytic dispersion of the SPhP $\omega_{\mathbf{k}}^{\mathrm{T}}$ in the local limit. As has previously been noted this is the dispersion relation arising from a Hamiltonian where a single photonic mode is coupled to a bath of LO phonons with coupling frequency \cite{Gubbin2021}
\begin{align}
	\lvert \Theta_{\mathbf{k}, n} \rvert^2 = \frac{\omega_{\mathrm{L}}^2 - \omega_{\mathbf{k}}^{\mathrm{T}\; 2}}{\omega_{\mathbf{k}}^{\mathrm{T}} \omega_{\mathbf{k}, n}^{\mathrm{L}}}  \frac{8 \beta_{\mathrm{L}}^2}{d^2}.
\end{align}
As discussed in Appendix D the full field of the polariton can be constructed from the separately quantised longitudinal and transverse fields of the LTP constituents by introduction of the weighting coefficients $\mathrm{T}_{\mathbf{k}}, \mathrm{L}_{\mathbf{k}}$. Then the total electric Hopfield field of the LTP is given by
\begin{equation}
    \boldsymbol{\theta}_{\mathbf{k}} = \mathrm{T}_\mathbf{k} \boldsymbol{\theta}_\mathbf{k}^{\mathrm{T}} +  \sum_n \mathrm{L}_{\mathbf{k}, n} \boldsymbol{\theta}_{\mathbf{k}, n}^{\mathrm{L}},
\end{equation}
where the transverse and longitudinal mode functions $\boldsymbol{\theta}_\mathbf{k}^{\mathrm{T}}, \boldsymbol{\theta}_{\mathbf{k}}^{\mathrm{L}}$ are defined and quantised individually in Appendix B and C respectively and the index $n$ runs over the discrete longitudinal phonon branches.

\subsection{Nonlocal Purcell Enhancement}
It is well known that emitters placed near to interfaces supporting plasmon polaritons can exhibit enhanced spontaneous emission due to the strong local field enhancement \cite{Barnes1998}. The enhancement can be described utilising the Purcell factor, which is the ratio of the emission rate of a dipole near to a photonic structure to that of the same dipole embedded in a homogeneous medium. It describes the enhancement in spontaneous emission induced by the resonator and is an especially interesting quantity in systems containing nanoscale features. In these systems predictions of large Purcell enhancements obtained with local optical theories can be modified when nonlocality is taken into account. Studies of plasmonic systems have demonstrated that the effect of nonlocality is to strongly diminish the achievable Purcell factor as a consequence of nonlocal quenching and transfer of energy out of the electromagnetic field into the kinetic energy of free electrons \cite{Tserkezis2016, Tserkezis2017}, possibly altering the emission regime \cite{Jurga2017, Tserkezis2018}. Nonlocality is expected to have a qualitatively different effect in polar systems, mostly because of the propagative nature of LO phonon modes within the Reststrahlen region. This allows energy transferred from an emitter into the system’s matter degrees of freedom to be coherently recycled as part of the LTP mode. Additionally, as the Purcell factor can be particularly strong in regions where the polariton group velocity is low, LTP systems could allow for enhanced spontaneous emission around their localised LO phonon frequencies, leading to large Purcell factors which can be tuned across the Reststrahlen region.
\begin{figure*}
\centering
\begin{subfigure}{.33\textwidth}
    \centering
    \includegraphics[width=\textwidth]{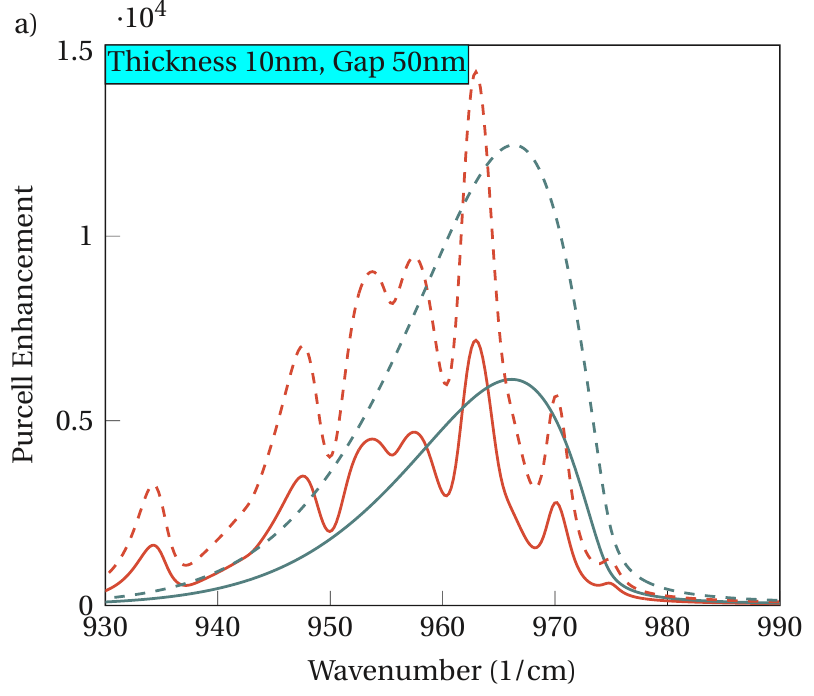}
\end{subfigure}%
\begin{subfigure}{.33\textwidth}
    \centering
    \includegraphics[width=\textwidth]{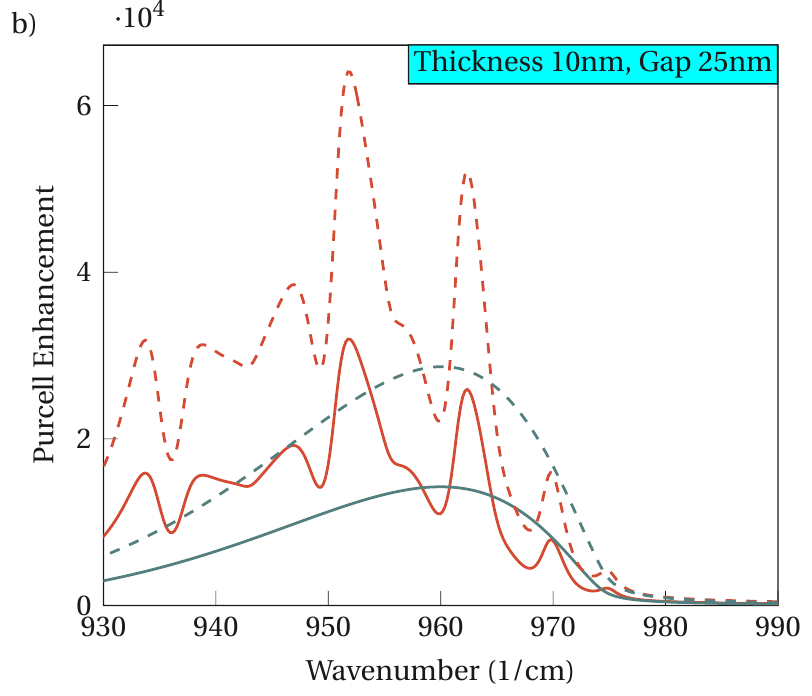}
\end{subfigure}%
\begin{subfigure}{.33\textwidth}
    \centering
    \includegraphics[width=\textwidth]{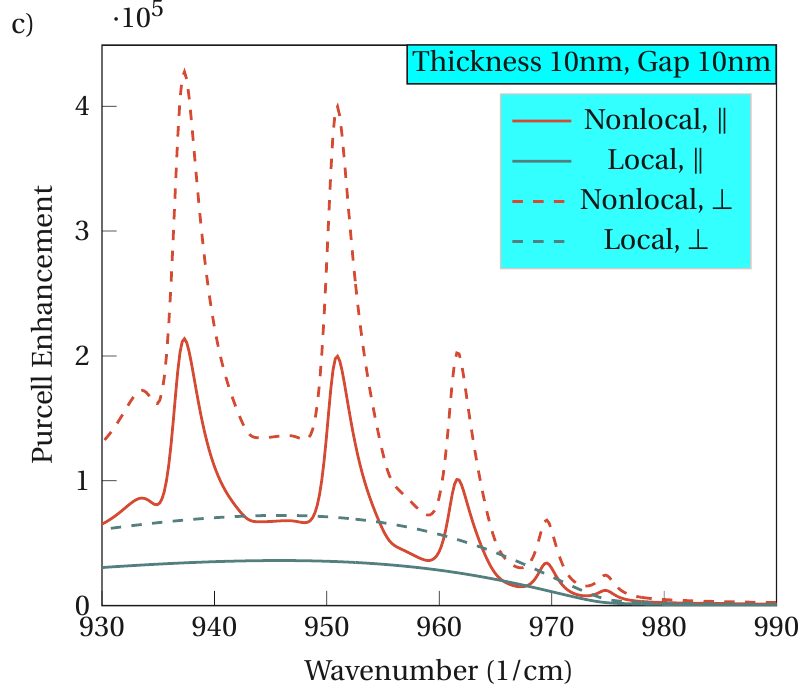}
\end{subfigure}%
\newline
\begin{subfigure}{.33\textwidth}
    \centering
    \includegraphics[width=\textwidth]{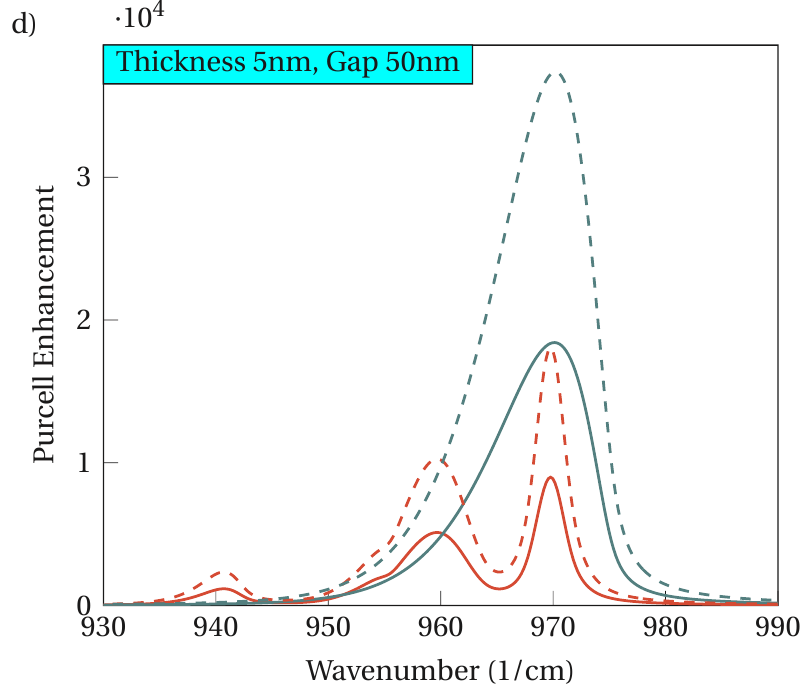}
\end{subfigure}%
\begin{subfigure}{.33\textwidth}
    \centering
    \includegraphics[width=\textwidth]{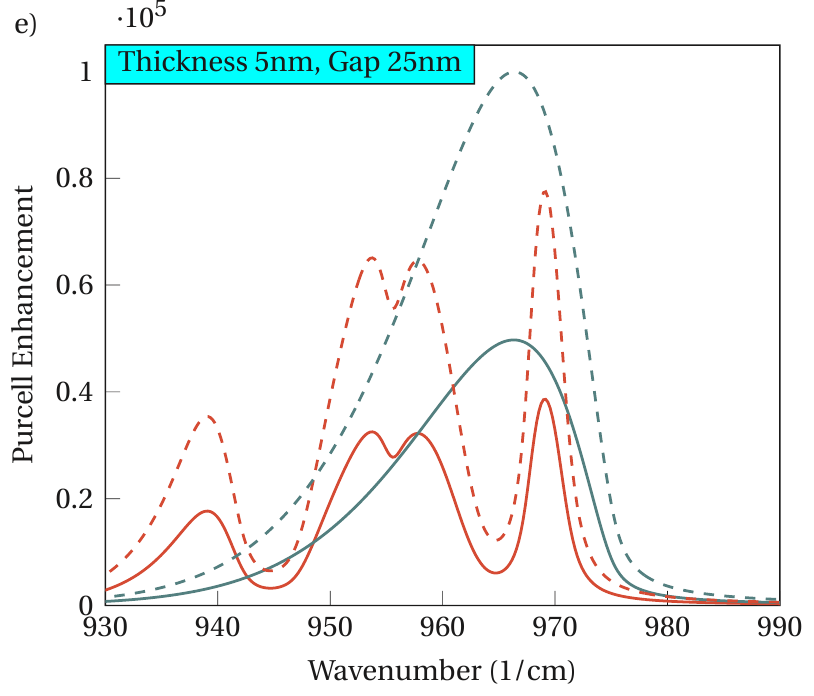}
\end{subfigure}%
\begin{subfigure}{.33\textwidth}
    \centering
    \includegraphics[width=\textwidth]{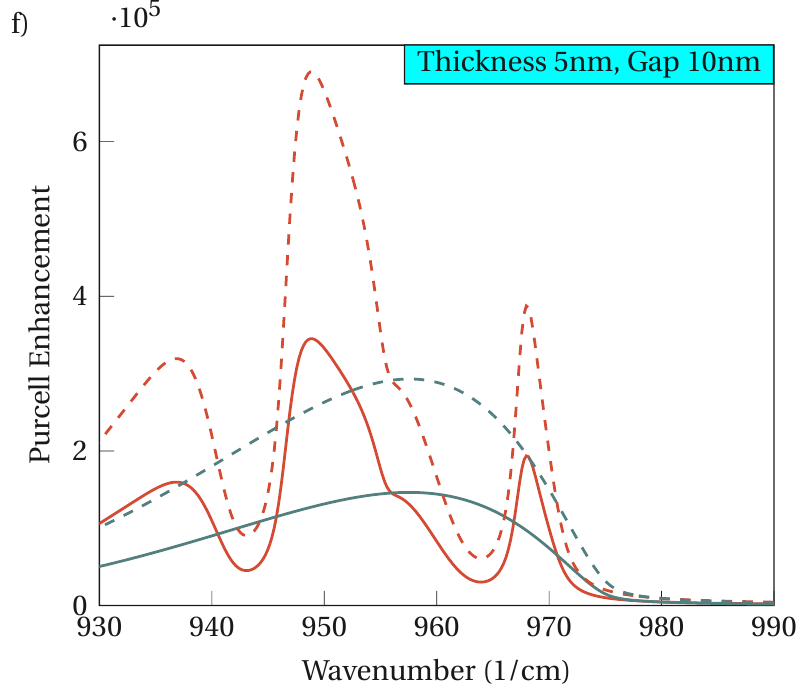}
\end{subfigure}%
\newline
\begin{subfigure}{.33\textwidth}
    \centering
    \includegraphics[width=\textwidth]{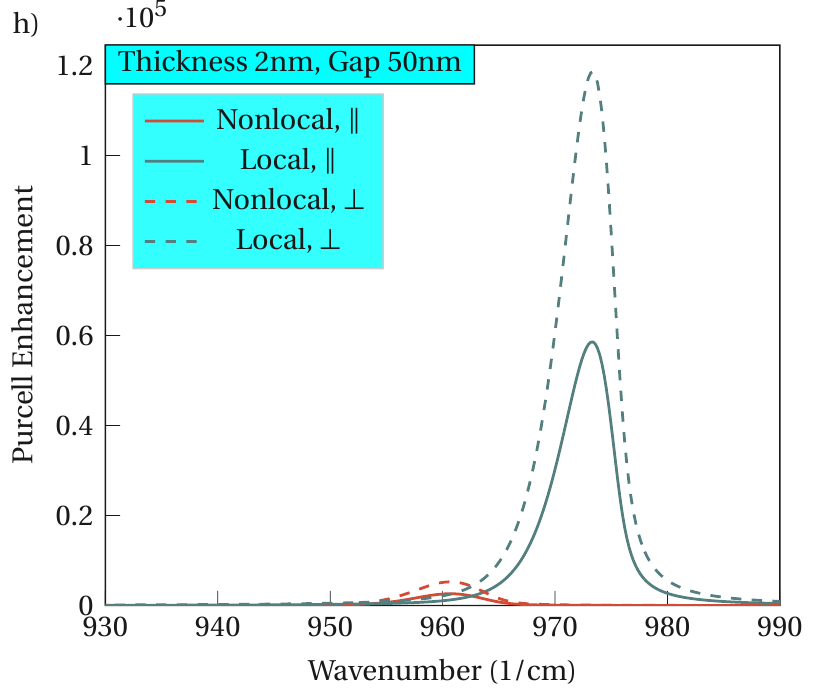}
\end{subfigure}%
\begin{subfigure}{.33\textwidth}
    \centering
    \includegraphics[width=\textwidth]{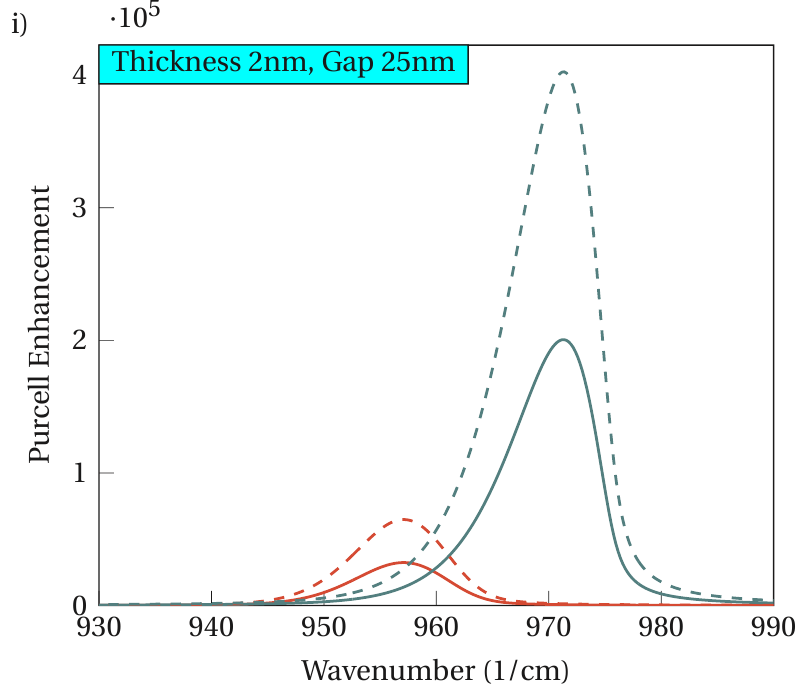}
\end{subfigure}%
\begin{subfigure}{.33\textwidth}
    \centering
    \includegraphics[width=\textwidth]{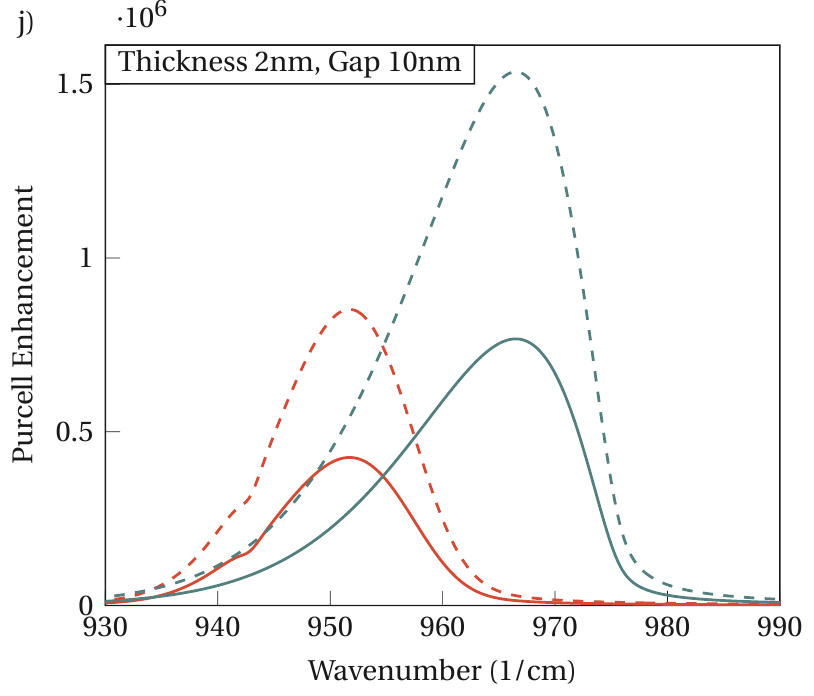}
\end{subfigure}%
	\caption{\label{fig:Fig2} Calculated Purcell enhancement for a 10nm (upper row), 5nm (middle row) and 2nm (bottom row) 3C-SiC films for a dipole located 50nm (left column), 25nm (middle column) and 10nm (right column) above the film. Green curves indicate the local enhancement, while red curves show the nonlocal. Enhancement for dipoles polarised parallel and perpendicular to the interface are indicated by solid and dashed lines respectively.}
\end{figure*}\\
To demonstrate this we apply our theory to the calculation of the Purcell enhancement for the system studied in the previous section. An example spectrum, calculated from the dispersion relation found in Appendix D is shown in Fig.~\ref{fig:Fig1} for a 10nm 3C-SiC film embedded between two high-index cladding regions. The epsilon-near-zero SPhP dispersion red-shifts with increasing wavevector \cite{Campione2015} undergoing avoided crossings \cite{Gubbin2016} with each coupled LO phonon branch \cite{Gubbin2021}. As each polariton branch heads toward it's asymptotic frequency its group velocity tends to zero.\\
In general we can write the quantised transverse electric field operator of the LTP as 
\begin{align}
	\hat{\mathbf{E}}_{\mathbf{k}} &=  \hbar \omega_{\mathbf{k}} \bar{\boldsymbol{\theta}}_{\mathbf{k}}^{\mathrm{T}} \hat{\mathcal{K}}_{\mathbf{k}} + \text{h.c} = \hbar \omega_{\mathbf{k}} \mathrm{T}_{\mathbf{k}} \mathrm{C}_{\mathbf{k}}  \mathbf{u}_{\mathbf{k}} \left(z\right) e^{i \mathbf{k} \cdot \mathbf{r} } \hat{\mathcal{K}}_{\mathbf{k} } + \mathrm{h.c.}, \label{eq:ESPhP}
\end{align}
where h.c. refers to the Hermitian conjugate, $\mathbf{k}$ is the wavevector in the waveguide plane, $\mathrm{T}_{\mathbf{k}}$ is the transverse weighting derived in Appendix D, $\mathrm{C}_{\mathbf{k}}$ is the quantisation constant derived in Appendix B, $\mathbf{u}_{\mathbf{k}}$ is the out of plane mode function derived in Appendix D and $\omega_{\mathbf{k}}$ is the LTP frequency derived from the nonlocal boundary conditions on the Hopfield equations in Appendix D.\\
We consider an emitter embedded in the upper positive dielectric halfspace, separated from the nonlocal film by a distance $h$. To derive meaningful results we need to include broadening for the LTP modes. Although this could have been achieved extending the nonlocal theory in the prior section through coupling the coherent system to a continuum of bath modes and Fano diagonalising the polariton system \cite{Gubbin2016b, Rajabali2021}, this is beyond the scope of this paper and is nevertheless not crucial due to the large quality factors of SPhPs and LTPs. Here we instead include losses phenomenologically through a Lorentzian density of states for the transition $\mathcal{X}$. The emission rate is given by
\begin{equation}
	\gamma = \frac{2 \pi}{\hbar} \sum_{\mathbf{k}} \mathcal{M}_{\mathbf{k}} \mathcal{X}\left(E_2 - E_1 - \hbar \omega_{\mathbf{k}}\right),
\end{equation}
in which the transition matrix element is given by
\begin{equation}
	\mathcal{M}_{\mathbf{k}} = \lvert \langle 1, n_{\mathbf{k}} + 1 \lvert \hat{\mathbf{d}} \cdot \hat{\mathbf{E}}_{\mathbf{k}} \rvert 2, n_{\mathbf{k}} \rangle \rvert^2,
\end{equation}
where $\hat{\mathbf{d}}$ is the dipole moment, states are labelled $\lvert i, j\rangle$ where $i$ labels the level of the dipole and $j$ the LTP system and $n_{\mathbf{k}}$ is the LTP population. Using Eq.~\ref{eq:ESPhP}, defining $\hat{\mathbf{d}}_{12} = \langle 1 \lvert \hat{\mathbf{d}} \rvert 2 \rangle$ and using the definition of the electric field Eq.~\ref{eq:ESPhP} we can write
\begin{equation}
	\mathcal{M}_{\mathbf{k}} = \frac{\hbar \omega_{\mathbf{k}} \lvert \mathrm{T}_{\mathbf{k}} \rvert^2}{4 \epsilon_0 \mathrm{L}_{\mathbf{k}}^{\mathrm{T}} \mathrm{S}} \lvert \mathbf{d}_{12} \cdot \mathbf{u}_{\mathbf{k}} \left(z\right) \rvert^2 \left(n_{\mathbf{k}} + 1 \right)
\end{equation}
Here the $n_{\mathbf{k}}$ term accounts for stimulated emission and the unity term for spontaneous emission. Ignoring the stimulated emission, we find the spontaneous emission rate for a dipole with transition frequency $\omega_0 = E_2 - E_1$ is given by
\begin{align}
	\gamma_{\mathrm{sp}} = \frac{2 \pi}{\hbar} \int_0^{\infty} \mathrm{d k} k & \frac{\mathrm{S}}{\left(2 \pi\right)^2} \frac{\hbar \omega_\mathbf{k} \lvert \mathrm{T}_{\mathbf{k}} \rvert^2}{4 \epsilon_0 \mathrm{S}} \mathcal{F}\left( E_2 - E_1 - \hbar \omega_\mathbf{k} \right) \nonumber \\
	& \int_0^{2 \pi} \mathrm{d} \theta \frac{\lvert \mathbf{d}_{12} \cdot \mathbf{u}_{\mathbf{k}} \left(z\right) \rvert^2}{\mathrm{L}_\mathbf{k}^{\mathrm{T}}}.
\end{align}
Evaluating the inner integral we obtain
\begin{align}
	\int_0^{2 \pi} \mathrm{d} \theta \frac{\lvert \mathbf{d}_{12} \cdot \mathbf{u}_{\mathbf{k}} \left(z\right) \rvert^2}{\mathrm{L}_\mathbf{k}^{\mathrm{T}}} &= \frac{2 \pi \lvert \mathbf{d}_{12} \rvert^2}{\mathcal{L}_{\mathbf{k}}^{\mathrm{T}} \left(z, \mathbf{d}_{12}, \omega_{\mathbf{k}}\right)},
\end{align}
where we defined unitless projections of the dipole moment operator $d_{12, \parallel}$ and $d_{12, z}$ using  $\mathbf{d}_{12, \parallel} \cdot \hat{\mathbf{k}} = d_{12} d_{12, \parallel} \cos \theta$ and defined $d_{12, z} = d_{12} \mathbf{d}_{12} \cdot \hat{\mathbf{z}}$, $\mathbf{d}_{12, \parallel} = \mathbf{d}_{12} - d_{12} \hat{\mathbf{z}}$ and the effective length
\begin{align}
	\frac{1}{\mathcal{L}_{\mathbf{k}}^{\mathrm{T}} \left(z, \mathbf{d}_{12}, \omega\right)} = \frac{e^{-2 \alpha_{\mathrm{C}} z}}{\lvert \epsilon_{\mathrm{C}} \rvert^2 \mathrm{L}_{\mathbf{k}}^{\mathrm{T}}} &\left[ \frac{1}{2} \lvert d_{12, \parallel} \rvert^2 \frac{\lvert \alpha_{\mathrm{C}} \rvert^2}{\lvert \mathbf{k} \rvert^2 }   +  \lvert d_{12, z}\rvert^2 \right] \nonumber \\
	&\times \biggr \lvert 1 + e^{- \alpha d} \biggr \rvert^2.
\end{align}
Dividing by the spontaneous emission rate of a dipole embedded in an infinite medium 
\begin{equation}
	\gamma_{\mathrm{sp}}^{0} = \frac{\omega_0^3 \lvert \mathbf{d}_{12} \rvert^2}{3 \pi \epsilon_0 \hbar c^3},
\end{equation}
we recover the Purcell factor
\begin{align}
	\mathrm{F}	&= \frac{\gamma_{\mathrm{sp}}}{\gamma_{\mathrm{sp}}^0}.
\end{align}

To demonstrate this result we calculate the Purcell factor for the thin-nonlocal film system for a few dipole-film separations. We have previously demonstrated that decreasing the thickness of a nonlocal film leads to a spectral spreading of the longitudinal mode frequencies away from the zone-centre longitudinal phonon, in conjunction with an increase in the LTP coupling strength as a result of the enhanced electromagnetic fields at the film interfaces \cite{Gubbin2021b}. The results are shown in Fig.~\ref{fig:Fig2}: where rows demarcate the different film thicknesses (10nm, 5nm, 2nm) and columns the for film-emitter separations (50nm, 25nm and 10nm). Emission rates for dipoles polarised perpendicular to the film are shown by dashed lines, while those parallel to the film are indicated by solid lines. In the local case (red curves) the Purcell enhancement exhibits a single peak for each combination of film thickness and separation, this represents the frequency where the photonic modes overlap with the emitter is strongest. The result is a peak because the field confinement increases with increasing in-plane wavevector (lower frequency), until at large enough in-plane wavevector the emitter no longer sits in the near-field of the mode.\\
In the nonlocal case (green curves) the Purcell enhancement breaks into distinct narrow peaks, these result from emission into the different polaritonic branches of the system. For the 10nm film there are many branches as the LO modes of the nanolayer are spectrally close. As the emitter-film separation decreases the number of peaks decreases and nonlocal emission outpaces the local result. This is because emitters very close to the nanolayer can emit into the low group-velocity flatband polariton regions where the density of final states is large. These flatband regions correspond to the peaks for 10nm separation. Similar physics is observed for the 5nm film in the middle row. There are less peaks in the nonlocal emission spectrum because the LO modes of the layer are more spaced out. For the 2nm film the LO modes are very far apart in frequency, and only one mode remains on the plot. As the emitter moves closer to the film the nonlocal peak red-shifts toward the flatband polariton region. This is a result of an enhanced density of states at the flatband polariton region, however there the fields become increasingly evanescent, only coupling to nearby emitters. This however happens more slowly as the LTP modes remain dispersive to larger in-plane wavevector for thinner layers \cite{Gubbin2020c}.\\
This plot shows that although the LTP is less photonic in nature ($\lvert \mathrm{T}_{\mathbf{k}_{\parallel}} \rvert^2 < 1$) than the local ENZ excitation it is able to enhance the Purcell factor. This is because near to the avoided crossing regions of the dispersion, and at large in-plane wavevectors the LTP group velocity drops enhancing the density of LTP states and leading to a peak in emission. In the regions between polariton branches emission is suppressed leading to a multi-peak Purcell spectra. In thinner films where coupling between LO phonons and SPhPs is enhanced this results in the narrowband emission enhancement.

\section*{Conclusion}
In this work we have presented a full quantum theory of longitudinal transverse polaritons, derived directly from consideration of the free fields in an inhomogeneous nonlocal medium. We derived the equations of motion for the Hopfield fields of the polariton, demonstrating them to be equivalent to the nonlocal macroscopic Maxwell equations shown previously. This point is important because it means that these equations can be solved by standard numerical means, allowing our theory to be integrated with the finite element or finite difference time domain solvers typically used to study SPhP resonators \cite{Gubbin2016b}. We also derived equations which allow for the quantisation of LTP modes. Finally we derived the nonlocal Purcell enhancement for a simple one-dimensional waveguide, demonstrating enhanced emission for dipoles close to the nonlocal layer. This is an example of how the tuneability of the LTP system could be exploited for enhanced light-matter interaction.
The quantum theory developed in this Paper is important for the study of novel and emergent phenomena in LTP systems. In a polar system nonlinear electron-electron scattering can be mediated by the charged ions comprising the lattice, leading to emission of LO phonons. If these LO phonons are part of an LTP this could permit direct electrical excitation or detection of SPhP modes, potentially underpinning a novel generation of mid-infrared optoelectronic devices \cite{Gubbin2022}. A quantum theory such as that presented here is necessary to calculate this emission rate. Similarly it can also be applied to descriptions of nonlinear LTP-LTP scattering mediated by anharmonic interactions between the underlying optical phonons, something which could allow for the formation of an LTP condensate analogous to the exciton-polariton condensates observed in the previous decade in the visible spectral region \cite{Daskalakis2014}.

\section*{Funding}
\label{SecAck}
S.D.L. is supported by a Royal Society Research fellowship and the  the Philip Leverhulme prize. The authors acknowledge support from the Royal Society Grant No. RGF\textbackslash EA\textbackslash181001 and the Leverhumlme Grant No. RPG-2019-174.

\section*{Author Declarations}
The authors have no conflicts to disclose.

\section*{Data Availability}
The data that support the findings of this study are available from the corresponding author upon reasonable request.

\bibliographystyle{naturemag}
\bibliography{bib}% Produces the bibliography via BibTeX.

\appendix

\section{Nonlocal Quantisation}
In this section we derive the nonlocal equations of motion for the Hopfield fields, given in the main body of the manuscript by Eq.~\ref{eq:Nleom1}-\ref{eq:Nleomn}. To achieve this we require the standard commutation relations between the canonical variables of the local Hamiltonian Eq.~\ref{eq:Hamloc} which are given by \cite{Gubbin2016b}
\begin{subequations}
	\begin{align}	
		\left[ \hat{\mathbf{D}}\left(\mathbf{r}\right), \hat{\mathbf{H}}\left(\mathbf{r}'\right)\right] &= \frac{i \hbar}{\mu_0}  \nabla' \times \delta\left(\mathbf{r}-\mathbf{r}'\right),\label{eq:comm2}\\
		\left[ \hat{\mathbf{X}} \left(\mathbf{r}\right), \hat{\mathbf{Q}}\left(\mathbf{r}'\right)\right] &= i \hbar \delta\left(\mathbf{r} - \mathbf{r}'\right),\label{eq:comm1}
	\end{align}
\end{subequations}
where the differential operator $\nabla'$ operates on $\mathbf{r}'$. These equations allow us to calculate the commutator of the polaritonic operator Eq.~\ref{eq:polaritondef} with the local Hamiltonian Eq.~\ref{eq:Hamloc}. As the commutation relation Eq.~\ref{eq:eom} is lengthy we proceed term by term, calculating the commutator of each polariton component with the Hamiltonian. Utilising the commutator between the electromagnetic fields Eq.~\ref{eq:comm2} we can find the following contributions to the polariton equation of motion Eq.~\ref{eq:eom}
\begin{widetext}
\begin{subequations}
\begin{align}
	\int \mathrm{d^3 r'}   \left[\boldsymbol{\alpha}\left(\mathbf{r}\right)\cdot\hat{\mathbf{D}}\left(\mathbf{r}\right), \frac{\mu_0 \hat{\mathrm{H}}\left(\mathbf{r}'\right)^2}{2}\right]  &=  i \hbar  \left[\nabla \times \boldsymbol{\alpha}\left(\mathbf{r}\right) \right] \cdot \hat{\mathbf{H}}\left(\mathbf{r}\right),\\
	\int \mathrm{d^3 r'}  \left[\boldsymbol{\beta}\left(\mathbf{r}\right) \cdot\hat{\mathbf{H}}\left(\mathbf{r}\right), \frac{\hat{\mathrm{D}}\left(\mathbf{r}'\right)^2}{2 \epsilon_0 \epsilon_{\infty}\left(\mathbf{r}'\right)},\right] &= - \frac{i \hbar c^2}{\epsilon_{\infty}\left(\mathbf{r}\right)}   \left[\nabla \times \boldsymbol{\beta}\left(\mathbf{r}\right)\right] \cdot \hat{\mathbf{D}}\left(\mathbf{r}\right),\\
	- \int \mathrm{d^3 r'} \frac{\kappa\left(\mathbf{r}\right)}{\epsilon_0 \epsilon_{\infty}\left(\mathbf{r}'\right)}  \left[\boldsymbol{\beta}\left(\mathbf{r}\right)\cdot\hat{\mathbf{H}}\left(\mathbf{r}\right), \hat{\mathbf{D}}\left(\mathbf{r}'\right)\cdot \hat{\mathbf{X}}\left(\mathbf{r}'\right)\right]  &=  i \hbar c^2  \frac{\kappa}{\epsilon_{\infty}} \left[\nabla \times \boldsymbol{\beta}\left(\mathbf{r}\right) \right] \cdot \hat{\mathbf{X}}\left(\mathbf{r}\right),
\end{align}
\end{subequations}
and furthermore through Eq.~\ref{eq:comm1} we can find
\begin{subequations}
\begin{align}
	\int \mathrm{d^3 r'} \frac{1}{2 \rho\left(\mathbf{r}'\right)} \left[\boldsymbol{\gamma} \left(\mathbf{r}\right)\cdot \hat{\mathbf{X}}\left(\mathbf{r}\right),  \hat{\mathrm{P}}\left(\mathbf{r}'\right)^2\right] &= \frac{i \hbar }{\rho\left(\mathbf{r}\right)}  \boldsymbol{\gamma}\left(\mathbf{r}\right) \cdot \hat{\mathbf{Q}}\left(\mathbf{r}\right) ,\\
	\int \mathrm{d^3 r'} \frac{\rho\left(\mathbf{r}'\right) \omega_{\mathrm{L}}\left(\mathbf{r}'\right)^2}{2} \left[\boldsymbol{\zeta}\left(\mathbf{r}\right) \cdot \hat{\mathbf{Q}}\left(\mathbf{r}\right),  \hat{\mathrm{X}}\left(\mathbf{r}'\right)^2\right] &= - i \hbar   \omega_{\mathrm{L}}\left(\mathbf{r}\right)^2 \rho\left(\mathbf{r}\right) \boldsymbol{\zeta} \left(\mathbf{r}\right) \cdot \hat{\mathbf{X}}\left(\mathbf{r}\right) ,\\
	- \int \mathrm{d^3 r'} \frac{\kappa\left(\mathbf{r}'\right)}{\epsilon_0 \epsilon_{\infty}\left(\mathbf{r}'\right)}  \left[\boldsymbol{\zeta}\left(\mathbf{r}\right) \cdot \hat{\mathbf{Q}}\left(\mathbf{r}\right),  \hat{\mathbf{D}}\left(\mathbf{r}'\right) \cdot \hat{\mathbf{X}}\left(\mathbf{r}'\right)\right] &= i \hbar \frac{\left[\kappa\left(\mathbf{r}\right)  \boldsymbol{\zeta}\left(\mathbf{r}\right) \right]^{\mathrm{T}}}{\epsilon_0 \epsilon_{\infty}\left(\mathbf{r}\right)} \cdot \hat{\mathbf{D}}\left(\mathbf{r}\right).
\end{align}
\end{subequations}
\end{widetext}
If we now collect terms proportional to the remaining operators to form Eq.~\ref{eq:eom} we find four local equations of motion
\begin{subequations}
	\begin{align}
		\omega \boldsymbol{\alpha} &= i  \frac{\left[\kappa  \boldsymbol{\zeta} \right]^{\mathrm{T}}}{\epsilon_0 \epsilon_{\infty}} - \frac{i  c^2}{\epsilon_{\infty}} \nabla \times \boldsymbol{\beta},  \label{eq:eomb1}\\
		\omega \boldsymbol{\beta} &= i  \nabla \times \boldsymbol{\alpha}, \\
		\omega \boldsymbol{\zeta} &= \frac{i}{\rho}  \boldsymbol{\gamma}, \label{eq:eomb3}\\
		\omega \boldsymbol{\gamma} &= - i  \omega_{\mathrm{L}}^2 \rho \boldsymbol{\zeta} + \frac{i  c^2 \kappa}{\epsilon_{\infty}} \nabla \times \boldsymbol{\beta}. \label{eq:eomb4}
	\end{align}
\end{subequations}
These equations can be solved by introduction of the novel electric Hopfield field $\boldsymbol{\theta}$
\begin{equation}
	\boldsymbol{\theta} = \boldsymbol{\alpha} +  \frac{\left[\kappa  \boldsymbol{\zeta} \right]^{\mathrm{L}}}{\epsilon_0 \epsilon_{\infty}}.
\end{equation}
The equation of motion for $\boldsymbol{\theta}$ can be derived combining the Eq.~\ref{eq:eomb3}, \ref{eq:eomb4} to find
\begin{equation}
	\left(\omega_{\mathrm{L}}^2 - \omega^2 \right) \boldsymbol{\zeta} = \frac{ c^2 \kappa}{\epsilon_{\infty} \rho} \nabla \times \boldsymbol{\beta},
\end{equation}
and eliminating $\boldsymbol{\zeta}$ from Eq.~\ref{eq:eomb1}
\begin{align}
	\omega \boldsymbol{\theta} &= \frac{i c^2}{\epsilon_{\infty}} \left[\frac{\omega_{\mathrm{L}}^2 - \omega_{\mathrm{T}}^2}{\omega_{\mathrm{L}}^2 - \omega^2}   - 1 \right] \nabla \times \boldsymbol{\beta} \nonumber \\
	&= - \frac{c^2}{\omega \epsilon_{\infty}} \left[\frac{\omega_{\mathrm{L}}^2 - \omega_{\mathrm{T}}^2}{\omega_{\mathrm{L}}^2 - \omega^2}   - 1 \right]  \nabla \times \nabla \times \boldsymbol{\theta}.
\end{align}
Finally on inversion this leads to
\begin{equation}
	\epsilon\left(\omega\right) \frac{\omega^2}{c^2} \boldsymbol{\theta} = \nabla \times \nabla\times \boldsymbol{\theta},
\end{equation}
where we recognised the local dielectric function of the polar lattice in the lossless limit
\begin{equation}
	\epsilon\left(\omega\right) =  \epsilon_{\infty} \frac{\omega_{\mathrm{L}}^2 - \omega^2}{\omega_{\mathrm{T}}^2 - \omega^2}.
\end{equation}

We can extend this theory to the nonlocal case, considering also commutation between the polaritonic operator $\hat{\mathcal{K}}$ and the nonlocal component of the Hamiltonian Eq.~\ref{eq:Hamfull}. To do this we need to derive additional commutation relationships between the ionic displacement field and it's spatial derivatives, which enter into the nonlocal component of the Hamiltonian. In the general case we can the commutator between momentum and the first derivative of the ionic position
\begin{align}
	&\int \mathrm{d^3 r'} \left[\hat{\mathrm{Q}}_i\left(\mathbf{r}\right),  \partial_m' \hat{\mathrm{X}}_l\left(\mathbf{r}'\right)   \right] \nonumber \\
	& \quad =  \int \mathrm{d^3 r'} \left[\hat{\mathrm{Q}}_i\left(\mathbf{r}\right),  \partial_m'  \right] \hat{\mathrm{X}}_l\left(\mathbf{r}'\right) + \partial_m' \left[\hat{\mathrm{Q}}_i\left(\mathbf{r}\right), \hat{\mathrm{X}}_l \left(\mathbf{r}'\right)  \right] \nonumber \\
	&\quad =  - i \hbar \delta_{i l} \partial_m' \delta\left(\mathbf{r} - \mathbf{r}'\right),  \label{eq:comm4}
\end{align}
where we noted that the derivative operator $\partial_m'$ operates on $\mathbf{r}'$ rather than $\mathbf{r}$ and therefore commutes with functions of $\mathbf{r}$. We also need to calculate commutation relations involving the stresses $\hat{\mathrm{S}}_{ij}$. Writing the nonlocal Hamiltonian density entering Eq.~\ref{eq:Hamfull} fully
\begin{equation}
	\hat{\mathcal{F}} = \frac{\bar{\mathcal{C}}_{ijkl} }{4} \left[ \frac{\partial \hat{\mathrm{X}}_i}{\partial x_j} + \frac{\partial \hat{\mathrm{X}}_j}{\partial x_i}\right]\left[ \frac{\partial \hat{\mathrm{X}}_k}{\partial x_l} + \frac{\partial \hat{\mathrm{X}}_l}{\partial x_k}\right],
\end{equation}
we commute
\begin{widetext}
\begin{align}
	&\int \mathrm{d^3 r}' \frac{\bar{\mathcal{C}}_{ijkl}' }{4} \left[\boldsymbol{\zeta \cdot \hat{\mathbf{Q}}}, \left( \frac{\partial \hat{\mathrm{X}}_i'}{\partial x_j'} + \frac{\partial \hat{\mathrm{X}}_j'}{\partial x_i'}\right)\left( \frac{\partial \hat{\mathrm{X}}_k'}{\partial x_l'} + \frac{\partial \hat{\mathrm{X}}_l'}{\partial x_k'}\right)\right] \nonumber \\
	& = \int \mathrm{d^3 r}' \frac{\bar{\mathcal{C}}_{ijkl}' }{4} \biggr\{  \left[\boldsymbol{\zeta \cdot \mathbf{Q}}, \frac{\partial \hat{\mathrm{X}}_i'}{\partial x_j'} + \frac{\partial \hat{\mathrm{X}}_j'}{\partial x_i'}\right] \left( \frac{\partial \hat{\mathrm{X}}_k'}{\partial x_l'} + \frac{\partial \hat{\mathrm{X}}_l'}{\partial x_k'}\right)  
	 +   \left( \frac{\partial \hat{\mathrm{X}}_i'}{\partial x_j'} + \frac{\partial \hat{\mathrm{X}}_j'}{\partial x_i'}\right) \left[\boldsymbol{\zeta \cdot \hat{\mathbf{Q}}},  \frac{\partial \hat{\mathrm{X}}_k'}{\partial x_l'} + \frac{\partial \hat{\mathrm{X}}_l'}{\partial x_k'}\right]  \biggr\}  \nonumber\\
	& =  \frac{i \hbar \bar{\mathcal{C}}_{ijkl} }{4} \biggr \{ \left( \frac{\partial^2 \hat{\mathrm{X}}_k}{\partial x_l \partial x_j} + \frac{\partial^2 \hat{\mathrm{X}}_l}{\partial x_k \partial x_j}\right) \zeta_i  + \left( \frac{\partial^2 \hat{\mathrm{X}}_k}{\partial x_l \partial x_i} + \frac{\partial^2 \hat{\mathrm{X}}_l}{\partial x_k \partial x_i}\right) \zeta_j +   \left( \frac{\partial^2 \hat{\mathrm{X}}_i}{\partial x_j \partial x_l} + \frac{\partial^2 \hat{\mathrm{X}}_j}{\partial x_i \partial x_l}\right) \zeta_k  + \left( \frac{\partial^2 \hat{\mathrm{X}}_i}{\partial x_j \partial x_k} + \frac{\partial^2 \hat{\mathrm{X}}_j}{\partial x_i \partial x_k}\right) \zeta_l \biggr \}, \label{eq:bigcommute}
\end{align}
\end{widetext}
in which we utilised the commutation relation Eq.~\ref{eq:comm4}
\begin{equation}
	\left[\sum_{y = m n o} \zeta_y \hat{\mathrm{Q}}_y, \frac{\partial \hat{\mathrm{X}}_i'}{\partial x_j'}\right]  =  - i \hbar \partial_j' \delta\left(\mathbf{r} - \mathbf{r}'\right) \sum_{y = mno} \zeta_y \delta_{y i},
\end{equation}
and the identity 
\begin{equation}
		\int \mathrm{d x} f\left(x\right) \delta'\left(x - x_0\right) = - f'\left(x_0\right).
\end{equation}
Note that in Eq.~\ref{eq:bigcommute} our underlying assumption of piecewise homogeneity prevented any material parameters contributing to the right hand side.

Unfortunately the derivatives in Eq.~\ref{eq:bigcommute} are on the field operators $\mathrm{X}$ rather than the Hopfield field $\zeta$ which precludes us from calculating the equation of motion. We can overcome this problem, collecting terms proportional to $\zeta_i$ to find the integrated quantity
\begin{align}
	&\int \mathrm{d^3 r}  \frac{i \hbar \bar{\mathcal{C}}_{ijkl} }{4}  \zeta_i \frac{\partial}{\partial x_j} \left( \frac{\partial \hat{\mathrm{X}}_k}{\partial x_l } + \frac{\partial \hat{\mathrm{X}}_l}{\partial x_k}\right) \nonumber \\
	&= - \int \mathrm{d^3 r}  \frac{i \hbar \bar{\mathcal{C}}_{ijkl} }{4 }   \frac{\partial \zeta_i}{\partial x_j} \left( \frac{\partial \hat{\mathrm{X}}_k}{\partial x_l } + \frac{\partial \hat{\mathrm{X}}_l}{\partial x_k}\right)  \nonumber \\
	&= \int \mathrm{d^3 r}  \frac{i \hbar \bar{\mathcal{C}}_{ijkl} }{4}  \left( \frac{\partial^2 \zeta_i}{\partial x_j \partial x_l} \hat{\mathrm{X}}_k + \frac{\partial^2 \zeta_i}{\partial x_j \partial x_k} \hat{\mathrm{X}}_l\right), 
\end{align}
where we assumed that the Hopfield field and it's derivatives vanish at the boundaries of the system. Similar results are derivable for the other terms in Eq.~\ref{eq:bigcommute}. Finally if we define the elements of the material stress tensor $\boldsymbol{\tau}$ as
\begin{equation}
	\tau_{ij} = \frac{\bar{\mathcal{C}}_{ijkl} \rho}{4} \left( \frac{\partial \zeta_k}{\partial x_l}  + \frac{\partial \zeta_l}{\partial x_k}\right),
\end{equation}
we can re-write the result of Eq.~\ref{eq:bigcommute}
\begin{widetext}
\begin{align}
	&\frac{i \hbar \bar{\mathcal{C}}_{ijkl} }{4} \biggr \{  \frac{\partial^2 \zeta_i}{\partial x_l \partial x_j} \hat{\mathrm{X}}_k + \frac{\partial^2 \zeta_i}{\partial x_k \partial x_j} \hat{\mathrm{X}}_l  +  \frac{\partial^2 \zeta_j}{\partial x_l \partial x_i} \hat{\mathrm{X}}_k + \frac{\partial^2 \zeta_j}{\partial x_k \partial x_i} \hat{\mathrm{X}}_l+    \frac{\partial^2 \zeta_k}{\partial x_j \partial x_l} \hat{\mathrm{X}}_i  + \frac{\partial^2 \zeta_k}{\partial x_i \partial x_l} \hat{\mathrm{X}}_j   +  \frac{\partial^2 \zeta_l}{\partial x_j \partial x_k} \hat{\mathrm{X}}_i + \frac{\partial^2 \zeta_l}{\partial x_i \partial x_k} \hat{\mathrm{X}}_j   \biggr \} \nonumber \\
	&= i \hbar \rho^{-1} \biggr \{ \frac{\partial \tau_{kl}}{\partial x_l} \hat{\mathrm{X}}_k  + \frac{\partial \tau_{kl}}{\partial x_k} \hat{\mathrm{X}}_l  +  \frac{\partial \tau_{ij}}{\partial x_j} \hat{\mathrm{X}}_i  +   \frac{\partial \tau_{ij}}{\partial x_i} \hat{\mathrm{X}}_j \biggr \} = i \hbar \rho^{-1} \biggr \{ \frac{\partial \tau_{lk}}{\partial x_l} \hat{\mathrm{X}}_k  + \frac{\partial \tau_{kl}}{\partial x_k} \hat{\mathrm{X}}_l  +  \frac{\partial \tau_{ji}}{\partial x_j} \hat{\mathrm{X}}_i  +   \frac{\partial \tau_{ij}}{\partial x_i} \hat{\mathrm{X}}_j \biggr \} \nonumber\\
	&= i \hbar \rho^{-1} \biggr \{ \left(\nabla\cdot \bar{\boldsymbol{\tau}}\right)_k \hat{\mathrm{X}}_k  + \left(\nabla\cdot \bar{\boldsymbol{\tau}}\right)_l \hat{\mathrm{X}}_l  +  \left(\nabla\cdot \bar{\boldsymbol{\tau}}\right)_i \hat{\mathrm{X}}_i  +   \left(\nabla\cdot \bar{\boldsymbol{\tau}}\right)_j \hat{\mathrm{X}}_j \biggr \} ,
\end{align}
\end{widetext}
where we noted the major symmetry of the stiffness tensor $\bar{\mathcal{C}}_{ijkl} = \bar{\mathcal{C}}_{klij}$, the minor symmetries of the stiffness tensor $\bar{\mathcal{C}}_{ijkl} = \bar{\mathcal{C}}_{jikl} = \bar{\mathcal{C}}_{ijlk}$ and in the final step utilised the tensor divergence
\begin{align}
	\left(\nabla\cdot \bar{\boldsymbol{\sigma}}\right)_i = \frac{\partial \sigma_{ki}}{\partial x_k}.
\end{align}
This modifies the fourth polaritonic equation of motion to
\begin{equation}
	\omega \boldsymbol{\gamma} = - i  \omega_{\mathrm{L}}^2 \rho \boldsymbol{\zeta} + \frac{i  c^2 \kappa}{\epsilon_{\infty}} \nabla \times \boldsymbol{\beta} + i \rho^{-1} \nabla \cdot \bar{\boldsymbol{\tau}},
\end{equation}
which is that utilised in the main body of the manuscript.

\section{Quantisation of the ENZ}
The epsilon-near-zero (ENZ) mode is a linear superposition of SPhPs at each interface of the thin film studied in the main body of the paper. The electric Hopfield field of the two SPhPs comprising the ENZ are defined in the main body of the Paper as
\begin{equation}
    \boldsymbol{\theta}_{u, \mathbf{k}} = \begin{cases}
    \frac{\alpha_\mathrm{C} \epsilon}{\alpha \epsilon_{\mathrm{C}}} \left[\frac{\mathbf{k}}{\lvert \mathbf{k} \rvert}, - \frac{i \lvert \mathbf{k} \rvert}{\alpha_\mathrm{C}} \right]^{\mathrm{T}}  e^{\alpha_\mathrm{C} \left(z + d\right)} e^{i \mathbf{k} \cdot \mathbf{r}} & z < -d,\\
	- \left[\frac{\mathbf{k}}{\lvert \mathbf{k} \rvert},  \frac{i \lvert \mathbf{k} \rvert}{\alpha}\right]^{\mathrm{T}} e^{-\alpha \left(z + d\right)} e^{i \mathbf{k} \cdot \mathbf{r}}  & -d < z < 0,\\
	- \frac{\alpha_\mathrm{C} \epsilon}{\alpha \epsilon_\mathrm{C}} \left[\frac{\mathbf{k}}{\lvert \mathbf{k} \rvert},  \frac{i \lvert \mathbf{k}\rvert}{\alpha_\mathrm{C}}\right]^{\mathrm{T}} e^{-\alpha_\mathrm{C} z - \alpha d} e^{i \mathbf{k} \cdot \mathbf{r}}  & z > 0,
	\end{cases}
\end{equation}      
\begin{equation}
    \boldsymbol{\theta}_{l, \mathbf{k}} = \begin{cases}
    \frac{\alpha_\mathrm{C} \epsilon}{\alpha \epsilon_\mathrm{C}} \left[\frac{\mathbf{k}}{\lvert \mathbf{k} \rvert}, - \frac{i \lvert \mathbf{k} \rvert}{\alpha_\mathrm{C}} \right]^{\mathrm{T}} e^{\alpha_\mathrm{C} \left(z + d\right) - \alpha d} e^{i \mathbf{k} \cdot \mathbf{r}} & z < -d,\\
	\left[\frac{\mathbf{k}}{\lvert \mathbf{k} \rvert},,  -\frac{i \lvert \mathbf{k} \rvert}{\alpha}\right]^{\mathrm{T}} e^{\alpha z } e^{i \mathbf{k} \cdot \mathbf{r}} & -d < z < 0,\\
	-\frac{\alpha_\mathrm{C} \epsilon}{\alpha \epsilon_\mathrm{C}} \left[\frac{\mathbf{k}}{\lvert \mathbf{k} \rvert},  \frac{i \lvert \mathbf{k} \rvert}{\alpha_\mathrm{C}}\right]^{\mathrm{T}} e^{-\alpha_\mathrm{C} z } e^{i \mathbf{k} \cdot \mathbf{r}} & z > 0,
	\end{cases}
\end{equation}
while the total transverse field is $ \boldsymbol{\theta}_{ \mathbf{k}}^{\mathrm{T}} =  \boldsymbol{\theta}_{l, \mathbf{k}} +  \boldsymbol{\theta}_{u, \mathbf{k}}$. The magnetic Hopfield field $\boldsymbol{\beta} = \frac{i}{\omega} \nabla \times \boldsymbol{\theta}$ can be written as
\begin{equation}
    \boldsymbol{\beta}_{u, \mathbf{k}} = \frac{i \omega \epsilon}{c^2 \alpha} \left[ \overrightarrow{\mathbf{z}} \times \frac{\mathbf{k}}{\lvert \mathbf{k} \rvert}, 0 \right]^{\mathrm{T}}  e^{i \mathbf{k} \cdot \mathbf{r}}  \begin{cases}
    e^{\alpha_\mathrm{C} \left(z + d\right)}& z < -d,\\
    e^{-\alpha \left(z + d\right)}  & -d < z < 0,\\
	e^{-\alpha_\mathrm{C} z - \alpha d}  & z > 0,
	\end{cases}
\end{equation}
\begin{equation}
    \boldsymbol{\beta}_{l, \mathbf{k}} =  \frac{i \omega \epsilon}{c^2 \alpha} \left[ \overrightarrow{\mathbf{z}} \times \frac{\mathbf{k}}{\lvert \mathbf{k} \rvert}, 0 \right]^{\mathrm{T}}  e^{i \mathbf{k} \cdot \mathbf{r}} \begin{cases}
    e^{\alpha_\mathrm{C} \left(z + d\right) - \alpha d}  & z < -d,\\
	e^{\alpha z } & -d < z < 0,\\
	e^{-\alpha_\mathrm{C} z } & z > 0,
	\end{cases}
\end{equation}
which as mentioned in the main body is continuous at material boundaries, while the tangential electric Hopfield field $\boldsymbol{\theta}$ is not. In the local case we would apply continuity of $\overrightarrow{\mathbf{z}} \times \boldsymbol{\theta}$ at the film interfaces, recovering the dispersion relation for odd-parity modes in the planar waveguide
\begin{equation}
    i \tan\left(\frac{\alpha d}{2}\right) = \frac{\epsilon \alpha_\mathrm{C}}{\epsilon_\mathrm{C} \alpha}.
\end{equation}

As we are interested in LTP modes we instead derive the quantisation condition in the general case. We start by integrating the fields in the barriers
\begin{multline}
    \int_{z \in \mathrm{C}} \mathrm{d} \mathbf{r}  \left\{ \frac{\hbar \omega}{\mu_0}  \lvert \beta_{\mathbf{k}}^{\mathrm{T}} \rvert^2 + \hbar \omega \epsilon_0 \epsilon_\mathrm{C}   \lvert \theta_{\mathbf{k}}^{\mathrm{T}} \rvert^2 \right\} \\
    = \frac{8 \epsilon_0 \mathrm{S} \lvert \epsilon \rvert^2 \hbar \omega}{\alpha^3 \epsilon_{\mathrm{C}} \Re\left\{\alpha_{\mathrm{C}}\right\}} \left(1 - d \Re\left\{\alpha \right\}\right) \left[i \alpha_{\mathrm{C}}^2 \Im\left\{\alpha\right\} + \lvert \mathbf{k} \rvert^2 \Re\left\{\alpha\right\}\right),
\end{multline}
where we expanded in powers of $d$ and retained terms to linear order. Here $\mathrm{S}$ is the in-plane quantisation surface. Finally we integrate over the film
\begin{multline}
    \int_{z \notin \mathrm{C}} \mathrm{d} \mathbf{r}  \left\{ \frac{\hbar \omega}{\mu_0}  \lvert \beta_{\mathbf{k}}^{\mathrm{T}} \rvert^2 + \hbar \omega \epsilon_0 \frac{\partial \left(\epsilon \omega\right)}{\partial \omega}   \lvert \theta_{\mathbf{k}}^{\mathrm{T}} \rvert^2 \right \} \\
    = 4 \epsilon_0 \mathrm{S}  \hbar \omega d \left[-  \frac{\alpha \epsilon^*}{\alpha^*} + \frac{\lvert \mathbf{k} \rvert^2}{\lvert \alpha \rvert^2} \left(\frac{\partial \left(\epsilon \omega\right)}{\partial \omega}  + \epsilon^*\right)\right]\\
    = \frac{4 \epsilon_0 \mathrm{S} \hbar \omega d}{\lvert \alpha \rvert^2} \left[-  \alpha^2 \epsilon^* + \lvert \mathbf{k} \rvert^2 \left(\omega \frac{\partial \epsilon}{\partial \omega}  + 2 \Re\left\{\epsilon^*\right\}\right)\right],
    \end{multline}
where again terms up to order $d$ were retained. In the lossless cladding $\Im\left\{\alpha_{\mathrm{C}} \right\} = 0,\; \Re\left\{\alpha_{\mathrm{C}} \right\} = \alpha_{\mathrm{C}}$ and similarly for the film so
\begin{equation}
    \frac{4  \epsilon_0 \mathrm{S} \hbar \omega \lvert \mathbf{k} \rvert^2 \epsilon^2}{\alpha^2} \left[\frac{2}{\epsilon_{\mathrm{C}} \alpha_{\mathrm{C}}} \left(1 - d \alpha \right) + d \left(-  \frac{\alpha^2}{\epsilon \lvert \mathbf{k} \rvert^2}  + \frac{1}{\epsilon^2} \frac{\partial \epsilon}{\partial \omega}  + \frac{2}{\epsilon} \right)\right],
\end{equation}
and if we define the quantisation length
\begin{equation}
    \mathrm{L}_{\mathbf{k}}^{\mathrm{T}} = \frac{2}{\epsilon_{\mathrm{C}} \alpha_{\mathrm{C}}} \left(1 - d \alpha \right) + d \left(-  \frac{\alpha^2}{\epsilon \lvert \mathbf{k} \rvert^2}  + \frac{\omega}{\epsilon^2} \frac{\partial \epsilon}{\partial \omega}  + \frac{2}{\epsilon} \right),
\end{equation}
we can write, for example, the quantised electric field coefficients
\begin{equation}
    \boldsymbol{\theta}_{u, \mathbf{k}} = \mathrm{C}_{\mathbf{k}} e^{i \mathbf{k} \cdot \mathbf{r}} \begin{cases}
    \frac{\alpha_\mathrm{C}}{\epsilon_{\mathrm{C}}} \left[\frac{\mathbf{k}}{\lvert \mathbf{k} \rvert^2}, - \frac{i}{\alpha_\mathrm{C}} \right]^{\mathrm{T}}  e^{\alpha_\mathrm{C} \left(z + d\right)}  & z < -d,\\
	-  \frac{\alpha}{\epsilon} \left[\frac{\mathbf{k}}{\lvert \mathbf{k} \rvert^2},  \frac{i}{\alpha}\right]^{\mathrm{T}} e^{-\alpha \left(z + d\right)}  & -d < z < 0,\\
	- \frac{\alpha_\mathrm{C}}{\epsilon_\mathrm{C}} \left[\frac{\mathbf{k}}{\lvert \mathbf{k} \rvert^2},  \frac{i}{\alpha_\mathrm{C}}\right]^{\mathrm{T}} e^{-\alpha_\mathrm{C} z - \alpha d}   & z > 0,
	\end{cases}
\end{equation}      
\begin{equation}
    \boldsymbol{\theta}_{l, \mathbf{k}} = \mathrm{C}_{\mathbf{k}} e^{i \mathbf{k} \cdot \mathbf{r}} \begin{cases}
    \frac{\alpha_\mathrm{C}}{\epsilon_\mathrm{C}} \left[\frac{\mathbf{k}}{\lvert \mathbf{k} \rvert^2}, - \frac{i}{\alpha_\mathrm{C}} \right]^{\mathrm{T}} e^{\alpha_\mathrm{C} \left(z + d\right) - \alpha d} & z < -d,\\
	\frac{\alpha}{\epsilon} \left[\frac{\mathbf{k}}{\lvert \mathbf{k} \rvert^2},  -\frac{i}{\alpha}\right]^{\mathrm{T}} e^{\alpha z }  & -d < z < 0,\\
	-\frac{\alpha_\mathrm{C}}{ \epsilon_\mathrm{C}} \left[\frac{\mathbf{k}}{\lvert \mathbf{k} \rvert^2},  \frac{i }{\alpha_\mathrm{C}}\right]^{\mathrm{T}} e^{-\alpha_\mathrm{C} z } & z > 0,
	\end{cases}
\end{equation}
where we defined
\begin{equation}
    \mathrm{C}_{\mathbf{k}} = \frac{1}{\sqrt{4 \epsilon_0 \hbar \omega \mathrm{L}_{\mathbf{k}}^{\mathrm{T}} \mathrm{S}}}.
\end{equation}
Note that in the main body of the text we denote the quantised transverse field by
\begin{equation}
    \boldsymbol{\theta}_{ \mathbf{k}}^{\mathrm{T}} =  \boldsymbol{\theta}_{l, \mathbf{k}} +  \boldsymbol{\theta}_{u, \mathbf{k}} = \mathrm{C}_{\mathbf{k}} e^{i \mathbf{k} \cdot \mathbf{r}} \mathbf{u}\left(z\right),
\end{equation}
where $\mathbf{u}\left(z\right)$ is the out-of-plane mode function of the excitation.

\section{Nonlocal Quantisation}
We consider a single longitudinal phonon branch in the central layer. The electric potential associated with LO phonons localised in the thin film $-d < z < 0$ is given by
\begin{equation}
	\phi_\mathbf{k}^{\mathrm{L}} = e^{i \mathbf{k} \cdot \mathbf{r}} \begin{cases}
	0 & z < - d, \\
	\sin\left[\xi \left(z+d/2\right)\right] & -d < z < 0, \\
	0 & z > 0,
	\end{cases}
\end{equation}
where $\mathbf{k}$ is the in-plane wavevector and $\xi$ is the out-of-plane wavevector. This relates to the electric field through $\mathbf{E}  = - \nabla \phi$. If the cladding layers act as hard barriers for the phonon the out-of-plane wavevector of the mode will be quantised, as for photons in a Fabry-P{\`e}rot cavity, characterised by integer index $n$ and out-of-plane wavevector $\xi_n = n \pi / d$. We hypothesise an an electric Hopfield field of form
\begin{widetext}
\begin{equation}
	\boldsymbol{\theta}_{\mathbf{k}, n}^{\mathrm{L}} = \mathrm{B}_{\mathbf{k}, n}^{\mathrm{L}} e^{i \mathbf{k} \cdot \mathbf{r}} \begin{cases} 
	0 & z < - d, \\
	\left[i\frac{\mathbf{k}}{\lvert \mathbf{k} \rvert} \sin\left(\xi_n \left(z+d/2\right)\right) , \frac{\xi_n}{\lvert \mathbf{k} \rvert} \cos\left(\xi_n \left(z+d/2\right)\right) \right]^{\mathrm{T}} & -d < z < 0, \\
	0 & z > 0,
	\end{cases}
\end{equation}
\end{widetext}
where $\mathrm{B}_{\mathbf{k}, n}$ is a normalisation constant to be determined from the quantisation conditions. As the phonon is entirely localised within region $2$ it can be quantised considering only the energy in the film yielding
\begin{align}
	\text{sgn} \left[\omega\right]	&= 2 \hbar \omega \epsilon_0 \epsilon_{\infty}  \frac{\omega^2 \left[\omega_{\mathrm{L}}^2 - \omega_{\mathrm{T}}^2\right]}{\left[\omega^2 - \omega_{\mathrm{T}}^2 + \beta_{\mathrm{L}}^2 (\lvert \mathbf{k} \rvert^2 + \xi_n^2)\right]^2} \int \mathrm{d^3 r} \lvert \boldsymbol{\theta}_{\mathbf{k}, n}^{\mathrm{L}} \rvert^2 \nonumber\\
	&= 2 \hbar \epsilon_0 \epsilon_{\rho}  \frac{ \omega^3  \omega_{\mathrm{L}}^2\left[1 - \omega_{\mathrm{T}}^2 / \omega_{\mathrm{L}}^2\right]^2}{\left[\omega^2 - \omega_{\mathrm{T}}^2 + \beta_{\mathrm{L}}^2 (\lvert \mathbf{k} \rvert^2 + \xi_n^2)\right]^2} \int \mathrm{d^3 r} \lvert \boldsymbol{\theta}_{\mathbf{k}, n}^{\mathrm{L}} \rvert^2.
\end{align}
Now we can utilise that for real $\xi$, valid for frequencies $\omega < \omega_{\mathrm{L}}$ and for odd values of $n$
\begin{align}
    \int \mathrm{d^3 r} \lvert \boldsymbol{\theta}_{\mathbf{k}, n}^{\mathrm{L}} \rvert^2 &
    = \frac{\mathrm{S}}{2 \lvert \mathbf{k} \rvert^2} \biggr[\left(\xi_n^2 - \lvert \mathbf{k} \rvert^2 \right)d \biggr]. \label{eq:inted}
\end{align}
This result allows us to identify a quantization factor similarly to that for the pure longitudinal modes on the LO phonon dispersion relation
\begin{align}
    \mathrm{B}_{\mathbf{k}, n}^{\mathrm{L}} &= \sqrt{\frac{\omega_{\mathrm{L}}^2}{\hbar \omega^3 \epsilon_0 \epsilon_{\rho} \mathrm{S} \mathrm{L}_{\mathbf{k}, n}^{\mathrm{L}}}},
\end{align}
where we defined the effective mode length as
\begin{align}
    \mathrm{L}_{\mathbf{k}, n}^{\mathrm{L}} &= d \biggr[\frac{\xi_n^2}{\lvert \mathbf{k} \rvert^2} - 1 \biggr] \frac{\left(\omega_{\mathrm{L}}^2 - \omega_{\mathrm{T}}^2\right)^2}{\left[\omega^2 - \omega_{\mathrm{T}}^2 + \beta_{\mathrm{L}}^2 \left(\lvert \mathbf{k} \rvert^2 + \xi_n^2\right)\right]^2}.
\end{align}
Here the first term collects the wavevector dependent components from the integral, and the latter accounts for the modal drift away from zone centre $\omega_{\mathrm{L}}$.

Finally we define a quantisation constant suitable for all non-quantised (continuous) values of $\xi$, given by
\begin{align}
    \mathcal{B}_{\mathbf{k}}^{\mathrm{L}} &= \sqrt{\frac{\omega_{\mathrm{L}}^2}{\hbar \omega^3 \epsilon_0 \epsilon_{\rho} \mathrm{S} \mathcal{L}_{\mathbf{k}}^{\mathrm{L}}}},
\end{align}
where
\begin{align}
    \mathcal{L}_{\mathbf{k}}^{\mathrm{L}} &= d \biggr[\frac{\xi^2}{\lvert \mathbf{k} \rvert^2} - 1 + \left(\frac{\xi^2}{\lvert \mathbf{k} \rvert^2} + 1\right) \frac{\sin\left(\xi d\right)}{\xi d} \biggr] \nonumber \\
    &\quad \quad \times \frac{\left(\omega_{\mathrm{L}}^2 - \omega_{\mathrm{T}}^2\right)^2}{\left[\omega^2 - \omega_{\mathrm{T}}^2 + \beta_{\mathrm{L}}^2 \left(\lvert \mathbf{k} \rvert^2 + \xi^2\right)\right]^2}.
\end{align}
We derived this quantity by carrying out the same calculation but not assuming that the integral over the film simplifies as in Eq.~\ref{eq:inted}. This is necessary to calculate the longitudinal and transverse Hopfield fields in Appendix D.

\section{Nonlocal Dispersion}
To uniquely determine the LTP dispersion relation we need to consider the longitudinal and transverse Hopfield fields together and apply the boundary conditions on them as discussed in Section I-C of the main body. In Appendix B we constructed the ENZ field so as it's magnetic field coefficient $\boldsymbol{\beta}$ is continuous. As the LO modes constructed in Appendix C are curl free they have no associated $\boldsymbol{\beta}$. We therefore need to construct the fields to satisfy the second Maxwell boundary condition on the parallel electric field coefficient $\boldsymbol{\theta} \times \overrightarrow{\mathbf{z}}$ and an additional nonlocal boundary condition. As discussed in the main body the appropriate choice of additional condition is on $\epsilon_{\infty} \boldsymbol{\theta} \cdot \overrightarrow{\mathbf{z}}$. The boundary condition on $\boldsymbol{\theta} \times \overrightarrow{\mathbf{z}}$ can be written
\begin{multline}
    i \mathcal{B}_{\mathbf{k}} \lvert \mathbf{k} \rvert \sin \left(\xi d / 2\right) \mathrm{L}_{\mathbf{k}} + \frac{\alpha}{\epsilon}  \left[1 - e^{- \alpha d}\right] \mathrm{C}_{\mathbf{k}} \mathrm{T}_{\mathbf{k}}  \\ 
    = - \frac{\alpha_{\mathrm{C}}}{\epsilon_{\mathrm{C}}} \left[1 + e^{- \alpha d}\right]\mathrm{C}_{\mathbf{k}} \mathrm{T}_{\mathbf{k}},
\end{multline}
where $\mathrm{L}_{\mathbf{k}}, \mathrm{T}_{\mathbf{k}}$ are expansion coefficients weighting the longitudinal and transverse components of the LTP. Note that in the transverse limit $\mathrm{L} \to 0$ this gives
\begin{equation}
    1 - \frac{\alpha \epsilon_{\mathrm{C}} - \alpha_{\mathrm{C}} \epsilon}{\alpha \epsilon_{\mathrm{C}} + \alpha_{\mathrm{C}} \epsilon} e^{-\alpha d} = 0,
\end{equation}
which is the standard dispersion relation for the ENZ mode in a symmetric trilayer waveguide \cite{Campione2015}. The boundary condition on $\epsilon_{\infty} \boldsymbol{\theta} \cdot \overrightarrow{\mathbf{z}}$ gives
\begin{multline}
    \epsilon_{\infty} \mathcal{B}_{\mathbf{k}} \frac{\xi}{\lvert \mathbf{k} \rvert} \cos \left(\xi d / 2\right) \mathrm{L}_{\mathbf{k}} - i \frac{\epsilon_{\infty}}{\epsilon} \left[1 + e^{- \alpha d}\right] \mathrm{C}_{\mathbf{k}} \mathrm{T}_{\mathbf{k}}  \\ 
    = - i \left[1 + e^{- \alpha d}\right]\mathrm{C}_{\mathbf{k}} \mathrm{T}_{\mathbf{k}}.
\end{multline}
Dividing through and defining $\xi' = i \xi$ we find the standard dispersion for symmetric modes in a trilayer nonlocal waveguide \cite{Gubbin2021b}
\begin{equation}
    \frac{\lvert \mathbf{k} \rvert^2}{\xi' \alpha} \left[1 - \frac{\epsilon}{\epsilon_{\infty} } \right] \tanh \left(\xi' d / 2\right) =  \left[\tanh\left(\alpha d / 2\right) + \frac{\epsilon \alpha_{\mathrm{C}}}{\alpha \epsilon_{\mathrm{C}}} \right],
\end{equation}
which is the form given in the main body. This equation uniquely determines the relationship between the modes frequency and wavevector. To find the transverse electric field we also need to determine the $\mathrm{L}_{\mathbf{k}}, \mathrm{T}_{\mathbf{k}}$ longitudinal-transverse expansion coefficients. This can be done using either equation, for example
\begin{multline}
    \mathrm{L}_{\mathbf{k}} = \frac{1}{\mathcal{B}_{\mathbf{k}} \sinh \left(\xi' d / 2\right)}  \\
    \times \left[ \frac{\alpha}{\epsilon} \left[1 - e^{- \alpha d}\right] + \frac{\alpha_{\mathrm{C}}}{\epsilon_{\mathrm{C}}}  \left[1 + e^{- \alpha d}\right] \right]  \frac{\mathrm{C}_{\mathbf{k}} \mathrm{T}_{\mathbf{k}}}{\lvert \mathbf{k} \rvert},
\end{multline}
and the normalisation condition
\begin{equation}
    1 = \lvert \mathrm{L}_{\mathbf{k}} \rvert^2 + \lvert \mathrm{T}_{\mathbf{k}} \rvert^2,
\end{equation}
derived in the main body under the assumption that the longitudinal and transverse fields are quantised. This gives
\begin{widetext}
\begin{align}
    \lvert \mathrm{T}_{\mathbf{k}} \rvert^2 &= 1 - \lvert \mathrm{L}_{\mathbf{k}} \rvert^2 = 1 - \frac{\lvert \mathrm{C}_{\mathbf{k}} \rvert^2}{\lvert \mathcal{B}_{\mathbf{k}} \rvert^2 \lvert \sinh \left(\xi' d / 2\right) \rvert^2} 
     \biggr \lvert \frac{\alpha}{\epsilon} \left[1 - e^{- \alpha d}\right] + \frac{\alpha_{\mathrm{C}}}{\epsilon_{\mathrm{C}}}  \left[1 + e^{- \alpha d}\right] \biggr \rvert^2  \frac{1}{\lvert \mathbf{k} \rvert^2} \lvert \mathrm{T}_{\mathbf{k}} \rvert^2,
\end{align}
\end{widetext}
which yields the expansion coefficients for the transverse field of the LTP, for example
\begin{equation}
    \boldsymbol{\theta}_{\mathbf{k}}^{\mathrm{T}} = \mathrm{T}_{\mathbf{k}} \left[\boldsymbol{\theta}_{l, \mathbf{k}} + \boldsymbol{\theta}_{u, \mathbf{k}}\right],
\end{equation}
where $\boldsymbol{\theta}_{l, \mathbf{k}}, \boldsymbol{\theta}_{u, \mathbf{k}}$ are the quantised transverse fields given by the final equations in Appendix B.

\end{document}